\documentclass[PAPER, cernpreprint, atlasdraft=false, texlive=2023, USenglish, texmf, orcidlogo]{atlasdoc}
 
\usepackage{atlaspackage}
\usepackage{atlasbiblatex}
\usepackage{multirow}
 
\usepackage[BSM=true,jetetmiss=true]{atlasphysics}

\graphicspath{{logos/}{figures/}}
 
\usepackage{ANA-SUSY-2020-16-PAPER-defs}

\AtlasTitle{Search for pair production of higgsinos in events with two Higgs bosons and missing transverse momentum in $\sqrt{s}=\SI{13}{\tev}$ \pp collisions at the ATLAS experiment}

\AtlasAbstract{
This paper presents a search for pair production of higgsinos, the supersymmetric partners of the Higgs bosons, in scenarios with gauge-mediated supersymmetry breaking. Each higgsino is assumed to decay into a Higgs boson and a nearly massless gravitino.
The search targets events where each Higgs boson decays into \bbbar, leading to a reconstructed final state with at least three energetic $b$-jets and missing transverse momentum. Two complementary analysis channels are used, with each channel specifically targeting either low or high values of the higgsino mass. The low-mass (high-mass) channel exploits \num{126} (\num{139}) \si{\ifb} of $\sqrt{s}=\SI{13}{\tev}$ data collected by the ATLAS detector during Run~2 of the Large Hadron Collider. No significant excess above the Standard Model prediction is found. At 95\% confidence level, masses between $\SI{130}{\gev}$ and $\SI{940}{\gev}$ are excluded for higgsinos decaying exclusively into Higgs bosons and gravitinos. Exclusion limits as a function of the higgsino decay branching ratio to a Higgs boson are also reported.
}

\PreprintIdNumber{CERN-EP-2023-278}

\AtlasJournal{Phys.\ Rev.\ D.}
\AtlasJournalRef{Phys. Rev. D 109 (2024) 112011}
\AtlasDOI{DOI:10.1103/PhysRevD.109.112011}

\hypersetup{pdftitle={ATLAS document},pdfauthor={The ATLAS Collaboration}}
 
\begin{document}
 
\maketitle

\section{Introduction}
\label{sec:intro}

Supersymmetry (SUSY)~\cite{Golfand:1971iw,Volkov:1973ix,Wess:1974tw,Wess:1974jb,Ferrara:1974pu,Salam:1974ig} is an extension of the Standard Model (SM) predicting the existence of a bosonic (fermionic) partner for each fermionic (bosonic) particle of the SM. If $R$-parity is conserved~\cite{Farrar:1978xj}, the lightest supersymmetric particle (LSP) cannot decay solely into lighter SM particles and consequently is stable, making it a potential dark matter candidate.
The Higgs boson \enquote{mass hierarchy problem} could be resolved by supersymmetry, with the divergent Higgs mass diagrams being canceled out by their SUSY counterparts~\cite{Sakai:1981gr,Dimopoulos:1981yj,Ibanez:1981yh,Dimopoulos:1981zb}. This class of \enquote{natural} SUSY models requires the superpartners of the top and bottom quarks (i.e., the stop \susy{t} and sbottom \susy{b}), the gluon (i.e., the gluino \susy{g}), and the bosons of an extended Higgs sector (i.e., the higgsinos \susy{H}) to be light~\cite{Barbieri:1987fn}. Such particles should be produced abundantly in proton--proton (\pp) collisions at the Large Hadron Collider (LHC). While the ATLAS and CMS collaborations have set strong limits on the masses of the gluino and the stop particles, the exclusion limits on the higgsino masses are much less stringent~\cite{SUSY-2017-02, CMS-SUS-20-004,SUSY-2018-02, SUSY-2018-41, SUSY-2018-05, ATLAS:2024zzz}, motivating searches for higgsinos at the LHC.
 
\begin{figure}[h]
\centering
\includegraphics[width=0.5\textwidth]{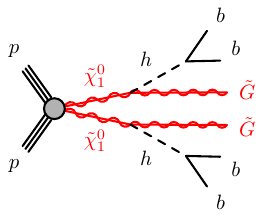}
\caption{Diagram of a signal event in the simplified SUSY model targeted by this analysis, where the \ninoone neutralino is treated as a pure higgsino.}
\label{fig:feyn_hhmodel}
\end{figure}

This paper presents a search for higgsino pair production inspired by models of general gauge mediation (GGM)~\cite{Meade:2008wd,Cheung:2007es,Dine:1981gu,AlvarezGaume:1981wy,Nappi:1982hm} or gauge-mediated supersymmetry breaking (GMSB)~\cite{Dimopoulos:1996vz,Matchev:1999ft}. These models are characterized by the existence of the lightest neutralino \ninoone, a neutral particle resulting from mixing between the SUSY partners of the SM electroweak bosons,
and a particle associated with spontaneous breaking of the global supersymmetry.
A range of phenomological scenarios are possible within this framework, depending on whether the \ninoone or the particle associated with supersymmetry breaking is the LSP. This paper studies the scenario in which SUSY is promoted to a local symmetry, as in many GMSB models. In this case the particle associated with supersymmetry breaking is absorbed by the superpartner of the graviton, the gravitino $\gravino$, and due to SUSY breaking being mediated at low energy, the gravitino is nearly massless. The \ninoone in these models is dominated by the higgsino component and treated as a pure higgsino. This allows the higgsino, which is produced via mass-degenerate pairs of charginos (the charged particles resulting from mixing between the SUSY partners of the SM electroweak bosons) or neutralinos, to decay into a SM Higgs boson and a nearly massless gravitino. The mass-degenerate production of \ninoone via both chargino and neutralino processes leads to an enhanced cross section compared to other models of electroweak SUSY production, making this an attractive process to search for discovery. The $\susy{H} \rightarrow h + \gravino$ mode dominates when $m_{\susy{H}}$ is greater than the Higgs mass and when $\tan\beta$ (the ratio of the vacuum expectation values of the Higgs doublets) is small~\cite{Meade:2009qv}. This specific scenario is implemented in this search through the simplified model represented in Fig.~\ref{fig:feyn_hhmodel}. The \gravino of the resulting model is effectively massless, with the mass set to $\SI{1}{\mev}$ for this analysis. The only free parameter in the model is the mass of the degenerate higgsino states, $m_{\susy{H}}$.
 
In this search, higgsinos are assumed to be produced in pairs, resulting in an experimental signature including two SM Higgs bosons and missing transverse momentum (\MET). Due to its high branching ratio, the \Hbb decay channel is an ideal target for this model, resulting in a final state with multiple $b$-jets (jets tagged as originating from $b$-quarks) and \MET.
The search is conducted in two complementary channels, each specifically targeting the production of either two high-mass or two low-mass higgsinos. Depending on the mass of the higgsino, the experimental signature can vary significantly, which motivates the use of different experimental approaches. The high-mass channel is characterized by significant \MET{} in the final state and relies on \MET-based triggers~\cite{TRIG-2019-01}. The low-mass channel employs a combination of \bjet triggers~\cite{TRIG-2018-08} because of the significantly lower \MET produced by low-mass higgsinos.
For a higgsino decay branching ratio of $\mathcal{B}(\susy{H}\rightarrow h + \gravino)=100\%$, the high-mass channel becomes the more sensitive one for $m_{\susy{H}}$ values above about $\SI{250}{\gev}$. The low-mass channel is used for results below this mass point, while the high-mass channel is used for results at all higher $m_{\susy{H}}$ values.
This strategy is used because the channels are not orthogonal and the sensitivity of the high-mass channel decreases rapidly as $m_{\susy{H}}$ falls below $\SI{250}{\gev}$. A similar search using the full \RunTwo dataset was performed by the CMS Collaboration~\cite{CMS-SUS-20-004}. Compared to the previous ATLAS search using 24.3${-}$36.1\,\ifb of \RunTwo data~\cite{SUSY-2017-02}, the present ATLAS search includes multiple improvements beyond the use of a larger dataset. These include a new method for pairing $b$-jets into Higgs boson candidates, significantly improved jet reconstruction and $b$-tagging, optimized $b$-tagging requirements in the low-mass channel, and improved  discrimination between signal and background in the high-mass channel through the use of multivariate techniques.
 
This paper is organized as follows. Section~\ref{sec:detector} describes the ATLAS detector, while Sec.~\ref{sec:data_mc_samples} describes the data and simulated samples. Section~\ref{sec:object_reco} defines the objects and other inputs used in reconstructing events for the two channels, Section~\ref{sec:event_sel_bkg} describes the event selection and background estimation, and Sec.~\ref{sec:systematics} the systematic uncertainties associated with this search. The results are reported in Sec.~\ref{sec:result}, while model-independent and model-dependent statistical interpretations are presented in Sec.~\ref{sec:stats}. The conclusions are given in Sec.~\ref{sec:conclusion}.


\section{ATLAS detector}
\label{sec:detector}

\newcommand{\AtlasCoordFootnote}{
ATLAS uses a right-handed coordinate system with its origin at the nominal interaction point (IP)
in the center of the detector and the \(z\)-axis along the beam pipe.
The \(x\)-axis points from the IP to the center of the LHC ring,
and the \(y\)-axis points upwards.
Cylindrical coordinates \((r,\phi)\) are used in the transverse plane,
\(\phi\) being the azimuthal angle around the \(z\)-axis.
The pseudorapidity is defined in terms of the polar angle \(\theta\) as \(\eta = -\ln \tan(\theta/2)\).
Angular distance is measured in units of \(\Delta R \equiv \sqrt{(\Delta\eta)^{2} + (\Delta\phi)^{2}}\).}
 
The ATLAS experiment~\cite{PERF-2007-01} at the LHC is a multipurpose particle detector
with a forward--backward symmetric cylindrical geometry and a near \(4\pi\) coverage in
solid angle.\footnote{\AtlasCoordFootnote}
It consists of an inner detector (ID) surrounded by a thin superconducting solenoid
providing a \SI{2}{\tesla} axial magnetic field, electromagnetic and hadron calorimeters, and a muon spectrometer.
The ID covers the pseudorapidity range \(|\eta| < 2.5\), and it consists of silicon pixel, silicon microstrip, and transition radiation tracking detectors.
Lead/liquid-argon (LAr) sampling calorimeters provide electromagnetic (EM) energy measurements
with high granularity.
A steel/scintillator-tile hadron calorimeter covers the central pseudorapidity range (\(|\eta| < 1.7\)).
The end cap and forward regions are instrumented with LAr calorimeters
for both the EM and hadronic energy measurements up to \(|\eta| = 4.9\).
The muon spectrometer (MS) surrounds the calorimeters and is based on
three large superconducting air-core toroidal magnets with eight coils each.
The field integral of the toroids ranges between \num{2.0} and \SI{6.0}{\tesla\metre}
across most of the detector.
The MS includes a system of precision chambers for tracking and fast detectors for triggering.
A two-level trigger system is used to select events.
The first-level trigger is implemented in hardware and uses a subset of the detector information
to accept events at a rate below \SI{100}{\kHz}.
This is followed by a software-based \enquote{high-level} trigger that
reduces the accepted event rate to \SI{1}{\kHz} on average
depending on the data-taking conditions.
An extensive software suite~\cite{ATL-SOFT-PUB-2021-001} is used in data simulation, in the reconstruction and analysis of real
and simulated data, in detector operations, and in the trigger and data acquisition systems of the experiment.


\section{Data and simulated samples}
\label{sec:data_mc_samples}

The data used in this search were collected by the ATLAS detector from \pp collisions produced during \RunTwo of the LHC, from 2015 to 2018. During this period, the LHC collided proton bunches at a center-of-mass energy of $\sqs = \SI{13}{\tev}$ with a bunch-crossing separation of \SI{25}{\nano\second}. The high-mass channel uses the complete \RunTwo dataset of \pp collisions, corresponding to an integrated luminosity of \SI{139}{\ifb} after requiring all detector subsystems to be operational and recording good quality data~\cite{DAPR-2018-01}. The data for this channel were collected with a combination of \met triggers.
The low-mass channel uses data collected through a combination of $b$-jet triggers, enabling it to focus on signal-like events with lower \met than in the high-mass channel, but with a lower integrated luminosity of \SI{126}{\ifb} due to some temporary operational issues associated with the $b$-jet trigger selections. The trigger selections, referred to as \enquote{online} selections, and the selections for fully reconstructed events, referred to as \enquote{offline} selections, are presented in Table~\ref{tab:lowmass_trig}.
The offline thresholds are tighter than the online ones in order to select events with a well-understood trigger efficiency. Additional offline selections are applied in the low-mass channel to provide independent event samples for each trigger employed. This approach allows each $b$-jet trigger to receive an independent efficiency correction for differences between data and Monte Carlo (MC) simulation.
 
\begin{table}[htb!]
\caption{Online and offline selections used for the high- and low-mass channels of the analysis. The second column, entitled \enquote{Year}, refers to the year the data were recorded. The $\met$ trigger requirement is listed in terms of $\met (\mu\;\text{inv.})$ because the trigger treats muons as being invisible~\cite{TRIG-2019-01}. The offline selections listed for the low-mass channel are only those required to ensure orthogonality between different trigger selections. The \HT variable corresponds to the scalar sum of the \pt of jets in the event. When a $\pt$ selection is listed for multiple jets, it is applied to each jet. The $p_{\text{T},j1}$ variable is the $\pt$ of the leading jet in an event.}
\label{tab:lowmass_trig}
\centering
{\small
\begin{tabular}[c]{l|c|>{\centering\arraybackslash}m{7cm}|c}
\toprule
Category & Year & Online selections & Offline selections \\
\midrule
\multicolumn{4}{c}{Low-mass channel} \\
\hline
\multirow{3}{*}{$2b1$j} & 2016 & 1 jet ($\pt > \SI{100}{\gev}$), \newline 2 $b$-jets ($60$\% $b$-jet efficiency, $\pt > \SI{55}{\gev}$)  & $p_{\text{T},j1} > \SI{150}{\gev}$ \\
\cline{2-4}  & 2017 & 1 jet ($\pt > \SI{150}{\gev}$), & $p_{\text{T},j1} > \SI{350}{\gev}$ \\
\cline{2-2} \cline{4-4} & 2018 &  2 $b$-jets ($70$\% $b$-jet efficiency, $\pt > \SI{55}{\gev}$) & $p_{\text{T},j1} > \SI{500}{\gev}$\\
\hline
\multirow{2}{*}{$2bH_\text{T}$} & 2017 & $\HT > \SI{300}{\gev}$, & $p_{\text{T},j1} < \SI{350}{\gev}$, $\HT > \SI{850}{\gev}$ \\
\cline{2-2} \cline{4-4} & 2018 & 2 $b$-jets ($50$\% $b$-jet efficiency, $\pt > \SI{55}{\gev}$) & $p_{\text{T},j1} < \SI{500}{\gev}$, $\HT > \SI{700}{\gev}$ \\
\hline
\multirow{6}{*}{$2b2$j} & \multirow{2}{*}{2016} & 2 jets ($\pt > \SI{35}{\gev}$), &  \multirow{2}{*}{$p_{\text{T},j1} < \SI{150}{\gev}$}\\
&  & 2 $b$-jets ($60$\% $b$-jet efficiency, $\pt > \SI{35}{\gev}$) &  \\
\cline{2-2}\cline{3-3}\cline{4-4} & \multirow{2}{*}{2017} & 2 jets ($\pt > \SI{35}{\gev}$), & \multirow{2}{*}{$p_{\text{T},j1} < \SI{350}{\gev}$, $\HT < \SI{850}{\gev}$}\\
&  &  2 $b$-jets ($40$\% $b$-jet efficiency, $\pt > \SI{35}{\gev}$)  &  \\
\cline{2-2}\cline{3-3}\cline{4-4} & \multirow{2}{*}{2018} & 2 jets ($\pt > \SI{35}{\gev}$),  & \multirow{2}{*}{$p_{\text{T},j1} < \SI{500}{\gev}$, $\HT < \SI{700}{\gev}$}\\
&  &  2 $b$-jets ($60$\% $b$-jet efficiency, $\pt > \SI{35}{\gev}$) &  \\
\midrule
\multicolumn{4}{c}{High-mass channel} \\
\hline
\multirow{4}{*}{\met} & 2015 & $\met (\mu\;\text{inv.}) > \SI{70}{\gev}$ & \multirow{4}{*}{$\met > \SI{150}{\gev}$} \\
\cline{2-3}
& 2016 & $\met (\mu\;\text{inv.}) > \SI{90}{\gev}$ & \\
\cline{2-3}
& 2017 & \phantom{0}$\met (\mu\;\text{inv.}) > \SI{100}{\gev}$ & \\
\cline{2-3}
& 2018 & \phantom{0}$\met (\mu\;\text{inv.}) > \SI{110}{\gev}$ &  \\
\bottomrule
\end{tabular}
}
\end{table}
 
MC simulations are used to model the signals and the high-mass channel background processes in this search. The signal samples consist of pair-produced higgsinos as described in Sec.~\ref{sec:intro}. These include both $\susy{H}\rightarrow h + \gravino$ and $\susy{H}\rightarrow Z + \gravino$ decays in order to test multiple values of the  $\susy{H}\rightarrow h + \gravino$ branching ratio. In addition, MC simulations of dijet processes are used to validate the background modeling for the high-mass channel. Table~\ref{t-MCsamples} shows the generator, set of tuned parameters (tune), parton distribution function (PDF) set, and cross section normalization order used for each sample. These samples were passed through a detailed simulation of the ATLAS detector and its response~\cite{SOFT-2010-01} based on \GEANT~\cite{Agostinelli:2002hh}.  A faster simulation relying on a parametrization of the calorimeter's response~\cite{ATL-PHYS-PUB-2010-013} is used to estimate the effect of the noninterference theory uncertainties in \ttbar and single-top production described in Sec.~\ref{sec:theory_syst}.
 
The effect of multiple interactions in the same and neighboring bunch
crossings (\pileup) was modeled by overlaying each simulated hard-scattering event with
inelastic \(pp\) events generated with \PYTHIA[8.186]~\cite{Sjostrand:2007gs}
using the \NNPDF[2.3lo] set of PDFs~\cite{Ball:2012cx} and the
A3 tune~\cite{ATL-PHYS-PUB-2016-017}. The MC events were weighted to reproduce the distribution of the average number of interactions per bunch crossing observed in the data. In all samples using the \PYTHIA parton shower model, the decays of bottom and charm hadrons were performed by \EVTGEN~\cite{Lange:2001uf}. In the generation of \ttbar events, the \hdamp parameter\footnote{The
\hdamp parameter is a resummation damping factor and one of the
parameters that controls the matching of \POWHEG matrix elements to
the parton shower and thus effectively regulates the
high-\pt radiation against which the \ttbar system recoils.} is set
to 1.5\,\mtop~\cite{ATL-PHYS-PUB-2016-020}. In samples produced with the \SHERPA generator, the matrix element calculations were matched
and merged with the \SHERPA parton shower based on Catani--Seymour
dipole factorization~\cite{Gleisberg:2008fv,Schumann:2007mg} using the \MEPSatNLO
prescription~\cite{Hoeche:2011fd,Hoeche:2012yf,Catani:2001cc,Hoeche:2009rj}.
The virtual QCD corrections were provided by the
\OPENLOOPS library~\cite{Buccioni:2019sur,Cascioli:2011va,Denner:2016kdg}. In the signal simulations, matrix elements for higgsino pairs were generated with up to two additional partons. Signal cross sections are calculated to next-to-leading order (NLO) in the strong coupling constant, adding the resummation of soft gluon emission at next-to-leading-logarithm accuracy (NLO+NLL)~\cite{Beenakker:1999xh,Bozzi:2007qr,Fuks:2012qx,Fuks:2013vua,Fuks:2013lya,Fiaschi:2018xdm}.
 
While simulations are used to estimate the high-mass channel backgrounds, the low-mass channel background is dominated by multijet processes that are not reliably modeled in simulation. A fully data-driven technique is used for background estimation in the low-mass channel as described in Sec.~\ref{sec:event_sel_bkg}.

\begin{table}[h]
 
\caption{List of generators used for the processes considered in this search, with MG5 standing for $\MADGRAPH$5. The
underlying-event (UE) tune, the PDF sets for the matrix element (ME) and UE, and the pQCD highest-order normalization accuracy used for each sample are also shown. Henceforth, the ${\ttbar}W$, ${\ttbar}Z$,
\tttt, and ${\ttbar}h$ processes are grouped into a single $\ttbar$+$X$ category. The SUSY signals consist of pair-produced higgsinos. The dijet samples are only used for validation.}
\label{t-MCsamples}
 
\centering{\footnotesize
\begin{tabular}{ccccc}
\toprule
Process &    Generator                   & Tune & PDF set&  Order of  \\
& + fragmentation/hadronization  &      &        &  cross section \\
\midrule{\normalfont \bfseries SUSY signals} &  MG5\_\AMCatNLO[2.6.1\,/\,2.6.2]~\cite{Alwall:2014hca}
& {\textsc A14}~\cite{ATL-PHYS-PUB-2014-021} & \NNPDF[2.3lo]~\cite{Ball:2012cx} & NNLO\textsubscript{approx}+NNLL~\cite{Beenakker:2016lwe,Beenakker:2014sma,Beenakker:2013mva,Beenakker:2011sf,Beenakker:2009ha,Kulesza:2009kq,Kulesza:2008jb,Beenakker:1996ch,Butterworth:2015oua} \\
& + \PYTHIA[8.230]~\cite{Sjostrand:2014zea} & & & \\
\\
{\normalfont \bfseries Dibosons}  & \SHERPA[2.2.1]~\cite{Bothmann:2019yzt,Gleisberg:2008fv,Buccioni:2019sur,Cascioli:2011va,Denner:2016kdg,Schumann:2007mg,Hoeche:2011fd,Hoeche:2012yf,Catani:2001cc,Hoeche:2009rj} & Default & \NNPDF[3.0nnlo]~\cite{Ball:2014uwa}  & NLO~\cite{Gleisberg:2008ta,Hoeche:2012yf} \\
{\normalfont \bfseries $WW$, $WZ$, $ZZ$} & & & & \\
\\
{\normalfont \bfseries $W$/${Z}+$jets} & \SHERPA[2.2.1] & Default & \NNPDF[3.0nnlo] & NNLO~\cite{Buccioni:2019sur,Cascioli:2011va,Denner:2016kdg}  \\
\\
{\normalfont \bfseries Top pairs: \ttbar} & \POWHEGBOX[v2]~\cite{Frixione:2007nw,Nason:2004rx,Frixione:2007vw,Alioli:2010xd}
& {\textsc A14} & \NNPDF[3.0nlo] (ME) & NNLO+NNLL~\cite{Beneke:2011mq,Cacciari:2011hy,Baernreuther:2012ws,Czakon:2012zr,Czakon:2012pz,Czakon:2013goa,Czakon:2011xx} \\
& + \PYTHIA[8.230] & & \NNPDF[2.3lo] (UE) & \\
\\
{\normalfont \bfseries Single top} & \POWHEGBOX[v2]
& {\textsc A14} & \NNPDF[3.0nlo] (ME) & NLO~\cite{Kant:2014oha,} ($t$/$s$-channel)\\
& + \PYTHIA[8.230] & & \NNPDF[2.3lo] (UE)&  NLO+NNLL~\cite{Kidonakis:2010ux,Kidonakis:2013zqa} ($Wt$)  \\
\\
{\normalfont \bfseries ${\ttbar}W$/${\ttbar}Z$} & MG5\_\AMCatNLO[2.3.3]~\cite{Alwall:2014hca}
& {\textsc A14} & \NNPDF[3.0nlo] (ME)& NLO~\cite{deFlorian:2016spz}\\
& + \PYTHIA[8.210] & &\NNPDF[2.3lo] (UE)&  \\
\\
{\normalfont \bfseries \tttt} & MG5\_\AMCatNLO[2.2.2]
& {\textsc A14} & \NNPDF[2.3lo] & NLO~\cite{Alwall:2014hca} \\
& + \PYTHIA[8.186] & & & \\
\\
{\normalfont \bfseries ${\ttbar}h$} & \POWHEGBOX[v2]
& {\textsc A14} & \NNPDF[3.0nlo] (ME) & NLO~\cite{deFlorian:2016spz} \\
& + \PYTHIA[8.230] & & \NNPDF[2.3lo] (UE)& \\
\\
{\normalfont \bfseries Dijet} & \PYTHIA[8.230] & A14 & \NNPDF[2.3lo] & LO \\
\bottomrule
\end{tabular}
}
\end{table}


\section{Object reconstruction}
\label{sec:object_reco}

Charged-particle tracks are required to have $\pt > \SI{0.5}{\gev}$. Primary vertex candidates are reconstructed from at least two charged-particle tracks~\cite{ATL-PHYS-PUB-2015-026}. To identify the hard-scattering interaction, the event's primary vertex is chosen as the vertex with the largest sum of squared track \pt ($\sum p_\text{T,track}^2$).
 
Small-radius jets are reconstructed using the \antikt algorithm~\cite{Cacciari:2008gp} with a radius parameter of $R=0.4$, with particle-flow objects as inputs. These objects are created by the particle-flow algorithm, which combines calorimeter energy clusters and ID tracks, improving the resolution of the combined energy measurement by subtracting the energy deposited by well-measured tracks in the calorimeter and using their \pT instead~\cite{PERF-2015-09}. Jets produced by collisions other than the hard scattering (i.e., \pileup jets) are removed by testing their compatibility with the primary vertex using the jet vertex tagger (JVT) discriminant~\cite{ATLAS-CONF-2014-018}. Jets with $\pt < \SI{60}{\gev}$ and $\abseta < 2.4$ are required to pass the \enquote{medium} JVT working point. Different MC-based calibration steps are applied to the reconstructed jets~\cite{JETM-2018-05}, including an area-based correction to account for energy contributions from \pileup interactions, a \pt- and $\eta$-dependent calibration to match the generator-level energy scale of the jets, and the \enquote{global sequential calibration} (GSC) to minimize energy calibration differences between quark- and gluon-initiated jets. Finally, an \textit{in situ} calibration is applied to jets in data to match the energy scale in simulation.
Sensitivity to signal scenarios where jets are close to each other due to large higgsino--gravitino mass splittings is enhanced by using large-radius jets produced by reclustering $R=0.4$ jets~\cite{ATLAS-CONF-2017-062} through another iteration of the \antikt algorithm with a radius parameter of $R=0.8$. For higgsinos with masses above $\SI{700}{\gev}$, over 50\% of events are expected to have at least one Higgs boson decay into a collimated cone of $R=0.8$. Calibrations are propagated through the iteration. The reclustered jets are trimmed~\cite{Krohn_2010} by removing any small-radius jets whose \pt falls below $f_\mathrm{cut}=$ 10\% of the \pt\ of the
large-radius jet. After this procedure, reclustered jets are required to have $\pt > \SI{100}{\gev}$ and $\abseta <2.0$.
The high-mass channel requires $R=0.4$ jets to have $\pt > \SI{25}{\gev}$ and $\abseta < 2.8$, while the low-mass channel uses jets with $\pt > \SI{40}{\gev}$ and $\abseta < 2.8$ because of requirements from its jet-based trigger strategy. The high-mass channel uses as a discriminant the total mass of the large-radius jets in the event, as explained later in Sec.~\ref{sec:HM_eventSel}.
 
Small-radius jets initiated by $b$-quarks and decaying within the ID acceptance ($\abseta < 2.5$) are identified as $b$-tagged jets using the \DLOner classifier set to a working point of 77\% efficiency for simulated \ttbar events~\cite{FTAG-2019-07}. This classification algorithm uses various types of inputs, including information about the impact parameters of ID tracks, the presence of displaced secondary vertices, and the reconstructed flight paths of $b$- and $c$-hadrons inside the jet. At the selected working point, the light-jet (charm-jet) rejection factor measured in \ttbar events is approximately 130 (4.9)~\cite{FTAG-2018-01,FTAG-2019-07,FTAG-2020-08,FTAG-2019-02}. Correction factors are applied to the simulated samples to account for differences in the $b$-tagging efficiencies between data and simulation. For the low-mass channel, additional correction factors are applied to account for differences in the online $b$-tagging efficiencies. Correlations between the trigger correction factors and offline correction factors are taken into account.
 
Three types of electrons and muons, \enquote{loose}, \enquote{baseline}, and \enquote{signal-quality}, are defined for this analysis. Electron candidates are built from energy deposits in the EM calorimeter that are matched to ID tracks~\cite{EGAM-2018-01}. Loose electrons are required to pass the \textit{LooseLH} likelihood identification criteria~\cite{PERF-2016-01} and to have $\pt > \SI{7}{\gev}$ and $\abseta < 2.47$. Further rejection of fake or nonprompt electrons is achieved by requiring electron tracks to match the primary vertex through a cut of $|z_0\sin\theta| < \SI{0.5}{\milli\meter}$ on the longitudinal impact parameter $z_0$. Loose electrons with $\pt > \SI{20}{\gev}$, called \enquote{baseline} electrons, are used to calculate the $\met$. To be considered as \enquote{signal-quality} objects, candidates are required to survive the overlap removal procedure defined below, satisfy the \textit{MediumLH} likelihood identification criteria~\cite{PERF-2016-01}, pass the \textit{Loose} isolation requirements, and have $\pt > \SI{20}{\gev}$ and $\abseta < 2.47$. Signal-quality electrons are also required to have a transverse impact parameter significance $|d_0|/\sigma(d_0) < 5$.
 
Muon candidates are reconstructed by matching an ID track with an MS track or performing a combined fit of an ID track with the aligned individual hits found in the MS. After reconstruction, loose muons are required to have $\pt > \SI{6}{\gev}$ and $\abseta < 2.7$, and to pass the \textit{medium} identification requirement based on track quality variables~\cite{MUON-2018-03}. Further rejection of fake or nonprompt muons is achieved by requiring muon tracks to match the primary vertex through a cut of $|z_0\sin\theta| < \SI{0.5}{\milli\meter}$ on the longitudinal impact parameter $z_0$ . Loose muons with $\pt > \SI{20}{\gev}$, called \enquote{baseline} muons, are used to calculate the $\met$ and to correct the four-momentum of jets to account for semileptonic $b$-hadron decays. This correction adds the muon four-momentum to the jet if a muon is found within $\Delta R =0.4$ of that jet. Signal-quality muons are the subset of loose muons that survive the overlap removal procedure defined below, pass the \textit{TightTrackOnly} (with variable radius) isolation requirements~\cite{MUON-2018-03}, and have $\pt > \SI{20}{\gev}$ and $\abseta < 2.5$. Signal-quality muons are also required to have a transverse impact parameter significance $|d_0|/\sigma(d_0) < 3$.
 
An overlap removal procedure is applied to resolve reconstruction ambiguities between electrons, muons, and small-radius jets. First, any baseline electron sharing an ID track with a baseline muon is rejected. Then, if a jet is found to be within $\Delta R = 0.2$ of a baseline electron, the jet is removed. If a baseline electron is found to lie $\Delta R < \min (0.4, 0.04 + \SI{10}{\gev}/\pt^e)$ from a remaining jet, where $\pt^e$ is the transverse momentum of the electron, the electron is removed. Next, any jet with an associated muon ID track or a baseline muon within $\Delta R = 0.2$ of its axis is removed if the jet has less than three tracks. Lastly, the muon is removed if it lies $\Delta R  < \min (0.4, 0.04 + \SI{10}{\gev}/\pt^\mu)$ from any remaining jet, where $\pt^\mu$ is the transverse momentum of the muon.
 
The missing transverse momentum \vecptmiss, with magnitude \met, is built from the negative vector sum of the transverse momenta of all well-identified and calibrated physics objects in the event, plus an extra \enquote{\met soft term}~\cite{ATLAS-CONF-2018-023} accounting for remaining low-energy charged particles. The \textit{tight} working point~\cite{ATLAS-CONF-2018-023} is used to reduce \pileup effects. The track-based soft term is calculated from ID tracks matched to the primary vertex but not to any physics object.  Baseline identification criteria, which are looser than the signal-quality criteria described above, are applied to muons and electrons used in these calculations.


\section{Event selection and background estimation}
\label{sec:event_sel_bkg}

Prior to any channel-specific selections, both channels impose data quality requirements to ensure that only events recorded when the entire ATLAS detector was fully operational are used~\cite{DAPR-2018-01}. These selections reject events containing corrupted data from the ID and calorimeters, as well as spurious jets caused by noncollision backgrounds~\cite{DAPR-2012-01,ATLAS-CONF-2015-029}.
Events containing signal-quality leptons (electrons or muons) are discarded to reduce backgrounds due to leptonically decaying $W$ bosons. Events are also discarded if they contain more than one loose lepton with $\pt > \SI{8}{\gev}$; a single loose lepton is allowed so as to avoid rejecting events containing a semileptonic $b$-hadron decay.
The loose-lepton criteria employed also minimize overlaps with ATLAS analyses using leptonic final states to target the same signal~\cite{SUSY-2018-02, SUSY-2018-05}.

\subsection{High-mass channel}

The high-mass channel focuses on detecting final states characterized by high \met{}, a minimum of three $b$-jets, and no signal-quality leptons. It relies on reconstructing the Higgs bosons resulting from the decay of higgsinos. To estimate the main backgrounds, MC simulations are used, with adjustments made to the normalization of \ttbar{} and $Z$+jets processes derived from data control samples. A boosted decision tree (BDT) is employed to distinguish between the signal and background events. A set of higgsino mass points, referred to as \enquote{mass hypotheses}, are considered, and for each of them control regions (CRs), validation regions (VRs), and up to four distinct signal regions (SRs) are defined by using the BDT score. Results for each mass hypothesis are obtained using a fit with only the corresponding regions defined for that hypothesis. The mass hypotheses are given by
\begin{equation*}
m_{\susy{H}} = \{200, 250, 300, 400, 500, 600, 700, 800, 900, 1000, 1100\}~\gev.
\end{equation*}
In addition, the three of these signal regions that have the highest expected sensitivity for the $\SI{250}{\gev}$, $\SI{500}{\gev}$, and $\SI{1000}{\gev}$ mass hypotheses, referred to as \enquote{discovery signal regions}, are used to search for, and set limits on, an excess of \enquote{beyond the Standard Model} (BSM) events in this phase space in a more model-independent manner.

\subsubsection{Event selection}\label{sec:HM_eventSel}
 
After the common selections described above and the trigger selections described in Sec.~\ref{sec:data_mc_samples}, signal-like events are required to satisfy the following \enquote{standard preselection} requirements:
\begin{itemize}
\item between four and seven small-radius jets with $\pt>\SI{25}{\gev}$ are reconstructed in the event, to reduce backgrounds with a large number of additional jets;
\item at least three of these jets are $b$-tagged according to the requirements described in Sec.~\ref{sec:object_reco}, as expected for the signal topology;
\item $\met$ is greater than $\SI{150}{\gev}$, for consistency with the production of invisible particles;
\item the minimum azimuthal angle between \met and any of the four leading jets (\dPhiMinFourJ) is greater than $0.4$, to reduce backgrounds with spuriously large \met resulting from mismeasurement of the momentum of a jet.
\end{itemize}
Scaling factors are applied to MC simulations to correct for discrepancies between simulated and data-based trigger efficiencies. Such corrections are negligible for events with $\met \ge \SI{200}{\gev}$ and reach a maximum of about 10\% for events with $\met \simeq \SI{150}{\gev}$ and a scalar sum of jet $\pt$ below $\SI{250}{\gev}$.
 
A key element of this analysis is the identification of the Higgs bosons originating from the higgsino decays. This is essential because the masses of the higher- and lower-mass Higgs boson candidates, denoted by \mHOne and \mHTwo, respectively, are used to discriminate between signal and background.
In order to obtain these values, the jets originating from the Higgs boson candidates must be identified and then paired. If there are exactly four $b$-jets in an event, those four are used. If there are more than four $b$-jets, the four with the highest \pt are used. If only three $b$-jets are reconstructed, and one of these jets has a mass greater than $\SI{100}{\gev}$, it is considered to be a boosted Higgs boson candidate and no additional jets are considered. Otherwise, the fourth jet is selected as the untagged small-radius jet that minimizes the value of \mHOne that would be obtained from the pairing algorithm discussed in the next paragraph.
 
The selected jets are paired to create Higgs boson candidates. If only three jets are selected (in the case where one jet has a mass greater than $\SI{100}{\gev}$), the heaviest jet is considered to be a Higgs boson candidate and the remaining two jets are paired to form the second candidate. Otherwise, the quantity $\dRbb_\text{max}(h_1^\text{HM},h_2^\text{HM}) = \max\left(\dR(h_1^\text{HM}),\dR(h_2^\text{HM})\right)$ is calculated for each of the three possible pairings of the four jets, where $\dR(h)$ is the $\dR$ separation of the jets coming from the same Higgs boson candidate. The pairing that minimizes $\dRbb_\text{max}$ is used, because pairs of jets have a more collimated topology in signal events than in background events.
 
In order to maximize the sensitivity to a broad set of higgsino masses, a BDT is used to discriminate between background and signal events and define the various regions for the high-mass channel. The BDT was trained, using the XGBoost algorithm~\cite{Chen:2016btl}, on inclusive background and signal datasets, properly reweighted to account for the different cross sections of the simulated processes. For the classification, the BDT exploited the following inputs:
\begin{itemize}
\item the number of jets \Nj and $b$-jets \Nbj in the event,
\item the scalar sum of the transverse momenta associated with small-radius jets in the event, $H_\text{T}$,
\item the magnitude \met of the missing transverse momentum, and the object-based \met significance, \metsig~\cite{ATLAS-CONF-2018-038},
\item the minimum azimuthal angle between \met and any of the four highest-\pt jets in the event, \dPhiMinFourJ,
\item the minimum transverse mass of the \met and the three leading $b$-jets in the event,
\begin{equation*}
\mtbmin =  \mathrm{min}_{i\leq 3}  \sqrt{(\met+\pT^{j_i})^2 - ({\met}_x+p_x^{j_i})^2 - ({\met}_y+p_y^{j_i})^2 },
\end{equation*}
\item the minimum angular distance between any two $b$-jets in the event, $\Delta R^{bb}_\mathrm{min}$,
\item the scalar sum of the masses of the reclustered large-radius jets in the event, \MJSigma,
\item the masses of the reconstructed Higgs boson candidates, \mHOne and \mHTwo, and the angular distances between the associated jets, $\dR(h_1^\text{HM})$ and $\dR(h_2^\text{HM})$.
\end{itemize}
The BDT was also parametrized with the generator-level higgsino mass \mHgen in order to define signal regions (SRs) that each target a specific mass hypothesis. The most discriminating variables for low-higgsino-mass hypotheses are \Nj, \Nbj, and \mtbmin, while \met, \metsig, and \mtbmin are the most discriminating variables for intermediate-to-high-higgsino-mass hypotheses. The optimization of the BDT hyperparameters was performed through a scan to optimize the discovery significance for the benchmark signal $m_{\susy{H}} = \SI{1000}{\gev}$. The selected parameters are 500 trees, a learning rate of 0.5, a maximum of 50 bins, and a maximum tree depth of six.
 
\begin{figure}[h!]
\centering
\subfloat[The SRs, CRs, and VRs targeting \ttbar and high \mtbmin events]{
\centering
\includegraphics[width=0.49\linewidth]{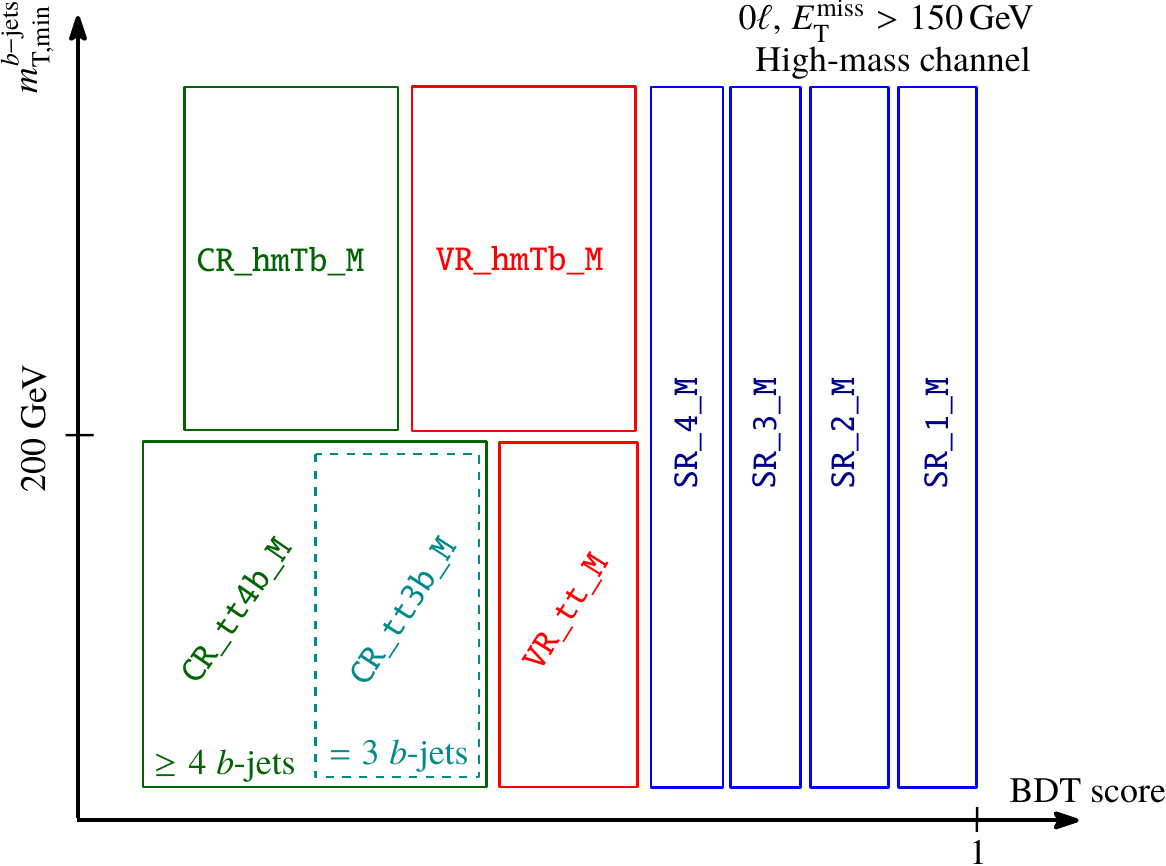}
\label{fig:RegionRecipe_SR}
}~
\subfloat[The CR and VR for $Z$+jets background]{
\centering
\includegraphics[width=0.49\linewidth]{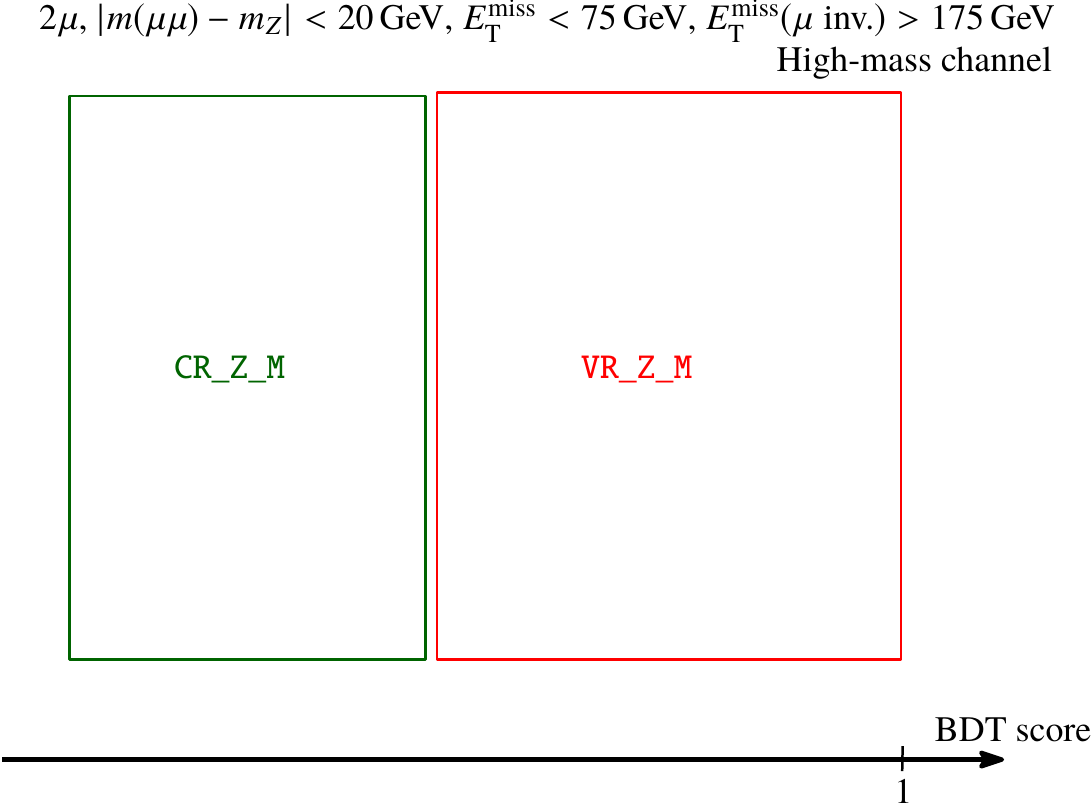}
\label{fig:RegionRecipe_Z}
}
\caption{General scheme followed in the definition of SRs, VRs, and CRs for each mass hypothesis of the high-mass channel. The left plot shows the SRs, \ttbar CRs and VR, and high \mtbmin CR and VR. The SRs are constructed with
all the events with BDT output scores in a specific range that maximizes the expected significance in the SR. Up to four SRs are defined that way. After
finding all SRs, the lowest BDT-score threshold of any SR for that specific mass hypothesis is used as the upper bound for the VRs, and the procedure is repeated separately in regions of low
and high \mtbmin to define VRs targeting the \ttbar background and high \mtbmin selections, as shown in the figure. Once the VRs are found, the process is repeated to define
CRs targeting \ttbar and high \mtbmin events, with a further splitting of the \ttbar CRs in $b$-jet multiplicity. The right plot shows the $Z$+jets CR and VR. These are defined using the same procedure as for the \ttbar and high \mtbmin regions, but with no SRs and a different preselection designed to capture $Z(\rightarrow \mu\mu)$+jets events. $\met (\mu\;\text{inv.})$ is used in place of $\met$ for the BDT input variables for the right plot to emulate the behavior of $Z(\rightarrow \nu\nu)$+jets. The BDT-score thresholds for the CRs are not shown in either plot; this is because they depend on the mass hypothesis and the type of CR.}
\label{fig:RegionRecipe}
\end{figure}

\begin{table}[h]
\centering
\caption{Summary of criteria applied to construct the CRs, VRs, and SRs of the high-mass channel. The considered variables are the statistical significance $Z$, the number of background events \nbkg, the ratio of signal events to background events \SoB, the number of $b$-jets \Nbj, and \mtbmin. The $Z$+jets preselection is discussed in Section~\ref{sec:bgr_High}. Additional selections based on the BDT score are included in the definitions of the CRs, VRs, and SRs, as shown in Figure~\ref{fig:RegionRecipe}; these are mass-dependent and are therefore omitted from the table.}
\label{table:SelectionsBDT}
 
\begin{tabular}{c|c|c|c|c|c|c}
\toprule
\multirow{2}{*}{Region name} & \multicolumn{3}{|c|}{Fixed Requirements} & \multicolumn{3}{c}{Boundary Conditions}\\
\cline{2-7}
& Preselection & \mtbmin & \Nbj & $Z$ & \nbkg & \SoB \\
\hline
\texttt{SR\_i\_M} & Standard & ... & ... & Maximize & \phantom{0.00}$\ge 0.5$ & ... \\
\texttt{VR\_tt\_M} & Standard & $<\SI{200}{\gev}$ & ... & ... & \phantom{0}$\ge 25$ & $<0.2$ \\
\texttt{VR\_hmTb\_M} & Standard & $>\SI{200}{\gev}$ & ... & ... & \phantom{0}$\ge 25$ & $<0.2$ \\
\texttt{VR\_Z\_M} & $Z$+jets & ... & ... & ... & \phantom{0}$\ge 25$ & $<0.2$ \\
\texttt{CR\_tt3b\_M} & Standard & $<\SI{200}{\gev}$ & $=3$ & ... & $\ge 100$ & $<0.1$ \\
\texttt{CR\_tt4b\_M} & Standard & $<\SI{200}{\gev}$ & $\ge 4$ & ... & $\ge 100$ & $<0.1$ \\
\texttt{CR\_hmTb\_M} & Standard & $>\SI{200}{\gev}$ & ... & ... & $\ge 100$ & $<0.1$ \\
\texttt{CR\_Z\_M} & $Z$+jets & ... & ... & ... & $\ge 100$ & $<0.1$ \\
\bottomrule
\end{tabular}
\end{table}
 
The SRs are defined using the BDT output scores of the preselected events. For each mass hypothesis, the SRs are built with an iterative procedure that begins with the highest BDT scores. The iterative process determines BDT-score thresholds for the SRs by requiring at least \num{0.5}~background events and maximizing the statistical significance $Z$ calculated from the \texttt{BinomialExpZ} function of RooStats~\cite{Moneta:2010pm}. For mass hypotheses that were excluded by the previous ATLAS search using 24.3${-}$36.1\,\ifb of \RunTwo data~\cite{SUSY-2017-02}, the cross sections for the signals are scaled down to the minimum previously excluded values for the purposes of this calculation. After an SR is formed, subsequent SRs are created by repeating the procedure on events with BDT scores below the values used in the previous SR. If this would result in a significance $Z$ of less than one, no SR is created and no further SRs are made for that mass hypothesis. If more than four SRs are created for a given mass hypothesis, the SRs with the lowest BDT scores are merged until there are only four SRs. The SRs are named as \texttt{SR\_i\_M}, where \texttt{M} corresponds to the signal mass hypothesis of the region and \texttt{i} is an integer between one and four that labels the SR, with lower values of \texttt{i} corresponding to SRs with higher-BDT-score requirements. Using different SRs for different signal mass hypotheses improves the sensitivity of the high-mass channel to low higgsino masses by approximately 20\% compared to using the same SRs for all mass hypotheses. The SRs corresponding to different signal mass hypotheses are not required to be orthogonal as the fit for each signal mass point only uses the regions defined for that mass hypothesis. These requirements are shown in Table~\ref{table:SelectionsBDT}, with a diagram of the SRs in Fig.~\ref{fig:RegionRecipe_SR}. The signal regions \texttt{SR\_1\_250}, \texttt{SR\_1\_500}, and \texttt{SR\_1\_1000} are additionally used to search for excesses with minimal model dependence and are called discovery regions when used in this context.
 
\subsubsection{Background estimation strategy} \label{sec:bgr_High}
 
The background estimation in the high-mass channel relies almost entirely on MC simulation with data-driven normalization corrections.
After preselection, the main background is $\ttbar$, followed by QCD multijet processes, primarily at low $\met$, as well as single top and $Z$+jets, which contribute at high $\met$. Smaller contributions arise from $\ttbar$+$X$ and diboson production. Distributions of the data and simulated backgrounds after preselection are shown in Fig.~\ref{fig:dataMC_HM}. The data and background predictions agree within 10\% after taking the statistical uncertainties into account. The normalizations of the dominant $\ttbar$ and $Z$+jets processes are measured for each signal mass hypothesis through a combined maximum-likelihood fit of the corresponding SRs and a dedicated set of control regions (CRs) with separately enhanced purity of each background component. Three additional parameters are included in each fit to constrain the normalizations of the single-top backgrounds, which make large but subleading contributions to the SRs, and the $\ttbar$+${\ge}1 b$ and $\ttbar$+${\ge}1c$ components of the $\ttbar$ background, to improve the modeling in heavy-flavor-dominated regions. The level of agreement between the adjusted background prediction and the data is checked through dedicated validation regions (VRs). These are named as \texttt{CR\_PROC\_M} and \texttt{VR\_PROC\_M}, where \texttt{PROC} labels the physical process and \texttt{M} denotes the signal mass hypothesis.
The \ttbar and high-$\mtbmin$ CRs and VRs are defined using the same selections as the SRs, except for the BDT score requirements. An additional requirement of $\mtbmin < \SI{200}{\gev}$ ($\mtbmin > \SI{200}{\gev}$) is used to enhance the purity of the \ttbar (high-$\mtbmin$) CRs and VRs. The \ttbar CRs are further split into $3b$ and $4b$ CRs to constrain the normalizations of the $\ttbar$+${\ge}1 b$ and $\ttbar$+${\ge}1c$ backgrounds separately from the overall $\ttbar$ background.
 
The $Z$+jets background is estimated using a separate data sample enriched in events containing muons. The dominant component of the $Z$+jets background in the SRs is $Z(\rightarrow \nu\nu)$+jets. The \enquote{$Z$+jets preselection} has the same requirements as the standard preselection described above, except for requiring two opposite-sign muons satisfying $\left|m(\mu\mu) - m_Z\right| < \SI{20}{\gev}$, where $m_Z$ is the mass of the $Z$ boson, and $\met < \SI{75}{\gev}$. The \met requirement is imposed to suppress the contamination from $\ttbar$ background and increase the sample's purity in $Z$+jets events. In addition to the \met requirement, the cut $\met (\mu\;\text{inv.}) > \SI{175}{\gev}$ is applied, where $\met (\mu\;\text{inv.})$ is computed with muons treated as invisible particles to emulate a boosted $Z\rightarrow \nu\nu$ decay. The $\met (\mu\;\text{inv.})$ requirement also allows events to be selected by the $\met$ trigger, which treats muons as being invisible~\cite{TRIG-2019-01}. Once these events are selected, the same BDT used for the signal selection is used to create the CRs and VRs of the $Z$+jets process. All BDT input variables using \met are adjusted to use $\met (\mu\;\text{inv.})$ instead.
 
For each signal mass hypothesis, the VR for a given background is defined to contain all events with BDT scores below the SR values and greater than a threshold value. The threshold BDT scores are selected such that each VR has at least a specific number of background events, denoted by $n_\textrm{bkg}$ in Table~\ref{table:SelectionsBDT}, and a signal contamination less than a threshold value, denoted by $S/B$ in Table~\ref{table:SelectionsBDT}. The CR is then defined using BDT scores below the values for the VR, with a minimum BDT score selected such that the CR has at least a specific number of background events and a signal contamination less than a threshold value, as indicated in Table~\ref{table:SelectionsBDT}. The strategy is illustrated graphically in Fig.~\ref{fig:RegionRecipe}. For the $Z$+jets CRs and VRs, $S/B$ is estimated to be zero and the signal contamination condition is therefore always satisfied. Due to their different preselection, the $Z$+jets VRs are not required to have BDT scores below the SR values.
 
A data-driven technique is used to estimate the QCD multijet background. This method exploits a template created by subtracting non-QCD-multijet backgrounds estimated with MC simulations from data in a kinematic regime dominated by QCD multijet processes. These events are obtained by replacing the selection $\dPhiMinFourJ > 0.4$  with $\dPhiMinFourJ < 0.2$. In order to estimate the contribution of this background to the CRs, VRs, and SRs, the template needs to be evaluated with the discriminating BDT. For this reason, a fake $\dPhiMinFourJ$ distribution is generated for the events in the template, sampled randomly from the expected $\dPhiMinFourJ$ distribution observed in dijet MC simulated samples. This template, however, does not correctly reproduce the correlations between $\dPhiMinFourJ$ and other kinematic variables that are used in the BDT training. Additionally, given the exponentially falling shape of the $\dPhiMinFourJ$ distribution for QCD multijet events, the template is not expected to accurately describe the normalization of this background in the targeted kinematic regime. These issues are resolved by applying a neural-network-assisted reweighting to the template. The neural network (NN) is trained on simulated dijet events to separate $\dPhiMinFourJ < 0.2$ events from $\dPhiMinFourJ > 0.4$ events, and an event weight based on the per-event NN output score is applied to the data-driven template to correct the normalization and the correlations between kinematic variables. The reweighted template is the final estimate for the QCD multijet background. This is validated in kinematic regimes enriched in QCD multijet events and is found to describe the data well, with up to 10\% differences. A nonclosure uncertainty of 50\% is applied to the QCD multijet prediction in the CRs, VRs, and SRs, found by comparing the performance of the NN-assisted reweighting with the prediction from dijet MC simulation in the analysis regions.

\begin{figure}[h!]
\centering
\subfloat[$\met$]{
\centering
\includegraphics[width=0.5\linewidth]{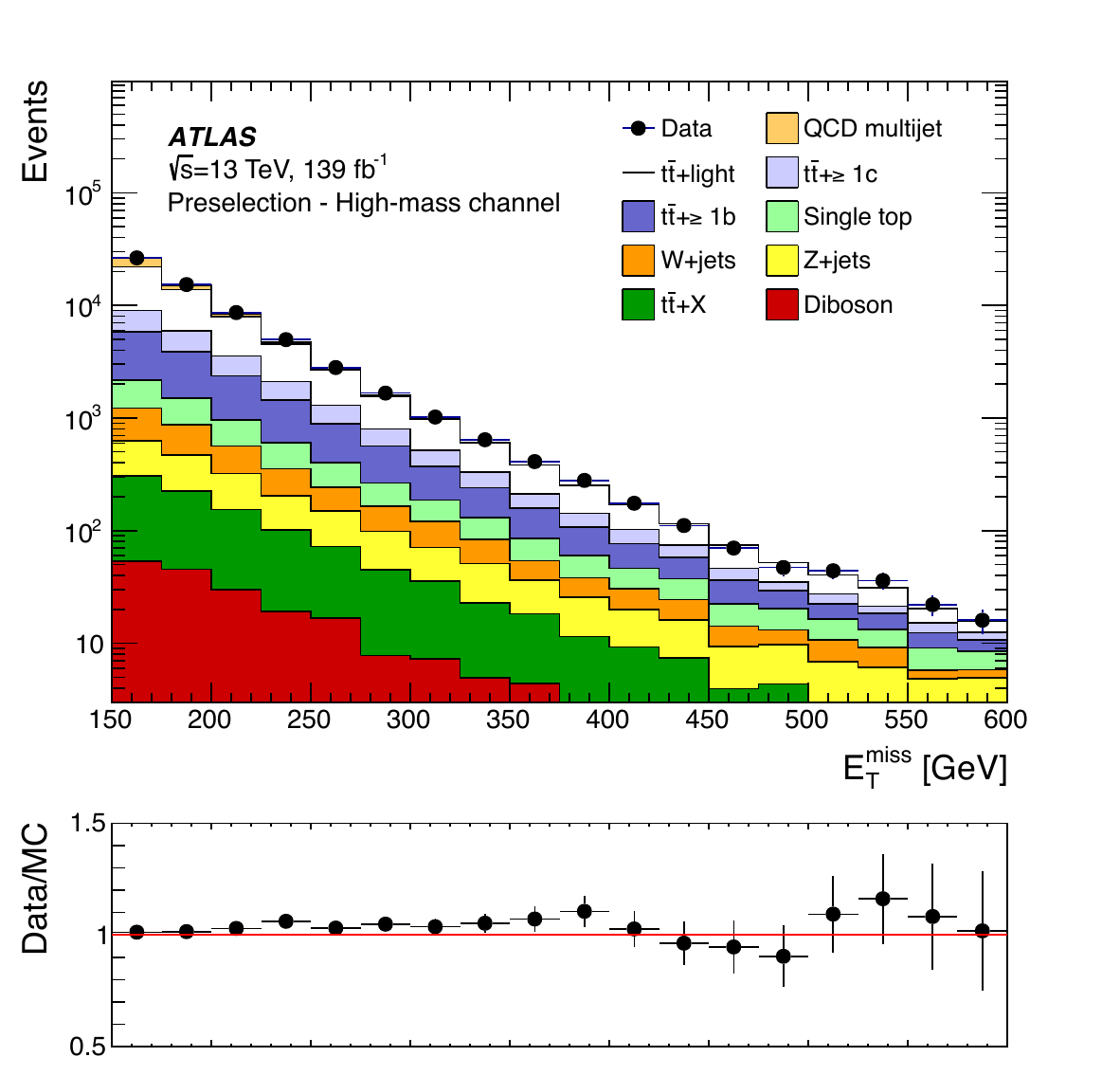}
\label{fig:dataMC_0L}
}~
\subfloat[$\mHOne$]{
\centering
\includegraphics[width=0.5\linewidth]{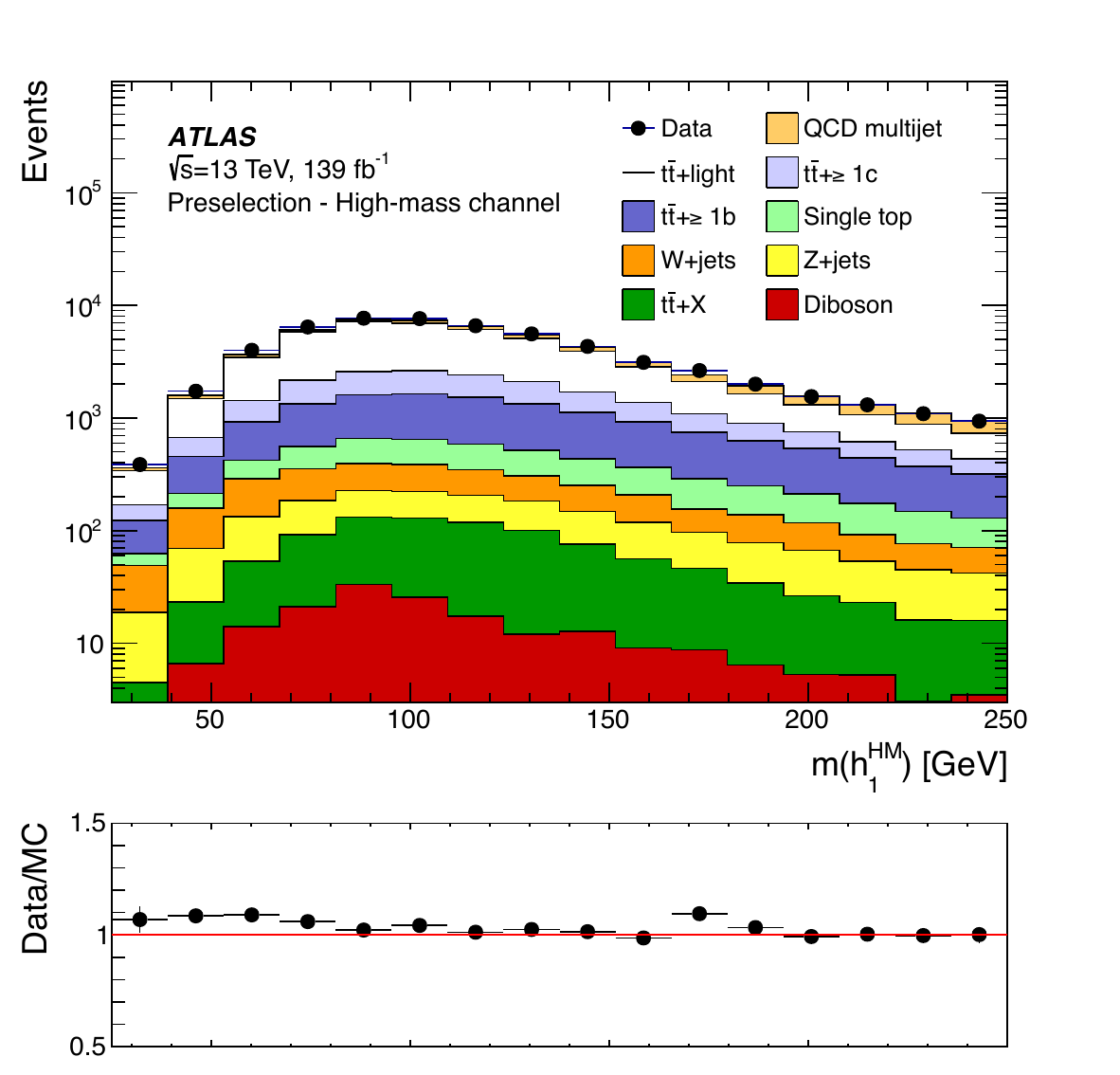}
\label{fig:dataMC_BDT500}
}
\caption{Comparisons of data and MC simulations for the standard preselection of the high-mass channel. The backgrounds are normalized to their theory cross sections. The left plot shows the distribution of $\met$ while the right plot shows the distribution of $\mHOne$. The lower panels show the ratios of data to MC simulations. The uncertainties in the ratios include the statistical uncertainties from both the data and MC simulations.}
\label{fig:dataMC_HM}
\end{figure}


\subsection{Low-mass channel}
 
The low-mass channel focuses on final states characterized by four or more $b$-jets and the absence of signal-quality leptons. Two Higgs boson candidates are reconstructed from the $b$-jets in the event. Signal, control, and validation regions are defined in the phase space of leading and subleading Higgs masses. In the low-mass channel, the primary sources of background are QCD multijet events and \ttbar{} processes. To estimate these backgrounds, a data-extrapolation method is employed, and a model is trained to adjust the kinematics across control regions with different $b$-jet multiplicities. The effectiveness of this adjustment is evaluated in validation regions. Within the signal region, two inclusive regions are defined in order to set model-independent limits on BSM physics. These regions are called \enquote{discovery regions} and are optimized to target higgsino mass hypotheses of $\SI{150}{\gev}$ and $\SI{300}{\gev}$, respectively.

\subsubsection{Event selection}
\label{sec:lm_ev}
 
The search in the low-mass channel uses events collected by the $b$-jet triggers listed in Table~\ref{tab:lowmass_trig}. Events are required to pass at least one of these triggers. Different offline kinematic requirements are also applied to ensure orthogonality between the different trigger selections, allowing correlations between the different trigger scale factors to be ignored.

Events are required to have at least four $b$-jets in order to reconstruct the Higgs boson candidates. These events are referred to as the \enquote{$4b$} sample. In events with more than four $b$-jets, only the four $b$-jets with the highest $\pt$ are considered. The four jets are then paired to form Higgs boson candidates. For each possible pairing, the quantity $\Delta R_\text{max}^{bb}(h_1^\text{LM},h_2^\text{LM})$ is calculated, where $h_1^\text{LM}$ is defined as the Higgs boson candidate with the higher reconstructed $\pt$ and $h_2^\text{LM}$ is defined as the one with lower $\pt$. The pairing that yields the smallest value of $\Delta R_\text{max}^{bb}$ is used.
 
The largest backgrounds passing these selections are QCD multijet and $\ttbar$ events. The leptonic $\ttbar$ background is reduced through the signal-quality and loose-lepton vetoes included in the common selections. Events are also required to have at most two loose leptons.\footnote{This veto uses all loose leptons, while the common loose-lepton veto only uses loose leptons with $\pt > \SI{8}{\gev}$.} In order to reduce the hadronic $\ttbar$ background, a discriminant based on the reconstruction of the top quark decay is used. Top quark candidates are reconstructed using three jets. One of these jets must be from a Higgs boson candidate and is considered to be the $b$-jet from the top quark decay. The other two jets are considered to form a $W$ boson candidate. Since a $W$ boson cannot decay into more than one $b$-jet, at least one of the $W$ boson candidate jets must not be associated with a Higgs boson candidate. The quantity
 
\begin{equation*}
\label{eqn:xwt}
X_{W\;\!\!t}=\sqrt{\left(\frac{m_{jj}-m_{W}}{0.1\cdot m_{jj}}\right)^2+\left(\frac{m_{jjb}-m_{t}}{0.1\cdot m_{jjb}}\right)^2}
\end{equation*}
 
is then calculated for each possible combination of jets, subject to the restrictions described above, and where $m_{jj}$ and $m_{jjb}$ are the masses of the reconstructed $W$ boson and top quark candidates, and $m_{W} = \SI{80.4}{\gev}$ and $m_{t} = \SI{172.5}{\gev}$ are the nominal masses of the $W$ boson and top quark. The factor of 0.1 approximates the fractional mass resolution for the reconstructed particle candidates. Events are vetoed if $X_{W\;\!\!t} < 1.8$ for any of these combinations.
 
The low-mass SRs are defined by the requirement $X_{hh}^\text{SR} < 1.6$, where $X_{hh}^\text{SR}$ is given by
 
\begin{equation*}
X_{hh}^\text{SR}=\sqrt{\left(\frac{m(h_1^\text{LM})-\SI{120}{\gev}}{0.1\cdot m(h_1^\text{LM})}\right)^2+\left(\frac{m(h_2^\text{LM})-\SI{110}{\gev}}{0.1\cdot m(h_2^\text{LM})}\right)^2},
\end{equation*}
 
where the denominators are the approximate mass resolutions for the Higgs boson candidates. A separate SR is created for each of the 2016, 2017, and 2018 data-taking periods to account for differences in the triggers and the background estimation procedure. The central values of $\SI{120}{\gev}$ and $\SI{110}{\gev}$ are offset from the true Higgs mass because of reconstruction inefficiencies such as from neutrinos produced in $b$-hadron semileptonic decays. The SR is the region within the innermost contour in Fig.~\ref{fig:lm_mass}.
 
The SRs are binned in $\met$ and $\meff$, where $\meff$ is defined as the $\met$ plus the scalar sum of the $p_\text{T}$ values of the jets associated with Higgs boson candidates. A fit is performed over the two-dimensional distribution of these variables, with lower bin edges given by
 
\begin{equation*}
\met = \{0, 20, 40, 60, 80, 100, 120, 140, 160, 180, 200\}~\gev
\end{equation*}
\begin{equation*}
\meff = \{160, 200, 260, 340, 440, 560, 700, 860\}~\gev,
\end{equation*}
 
where the last bin is inclusive. In addition, two discovery regions are defined in order to obtain limits in model-independent scenarios. These regions are optimized for the $m_{\susy{H}}=\SI{150}{\gev}$ (\texttt{SR\_LM\_150}) and $m_{\susy{H}}=\SI{300}{\gev}$ (\texttt{SR\_LM\_300}) mass points, although they are also sensitive at nearby $m_{\susy{H}}$ values. The region definitions are shown in Table~\ref{tab:lowmass_disc}.
 
\begin{table}[htb!]
\centering
\caption{Discovery region definitions for the low-mass channel.}
\label{tab:lowmass_disc}
\begin{tabular}[c]{l|c|c}
\toprule
Region & $\met$ & $\meff$ \\
\midrule
\texttt{SR\_LM\_150} & \phantom{0}$ > \SI{20}{\gev}$ & $> \SI{560}{\gev}$\\
\hline
\texttt{SR\_LM\_300} & $ > \SI{150}{\gev}$ & $> \SI{340}{\gev}$\\
\bottomrule
\end{tabular}
\end{table}
 
\begin{figure}
\centering
\subfloat[$4b$ data]{
\centering
\includegraphics[width=0.5\linewidth]{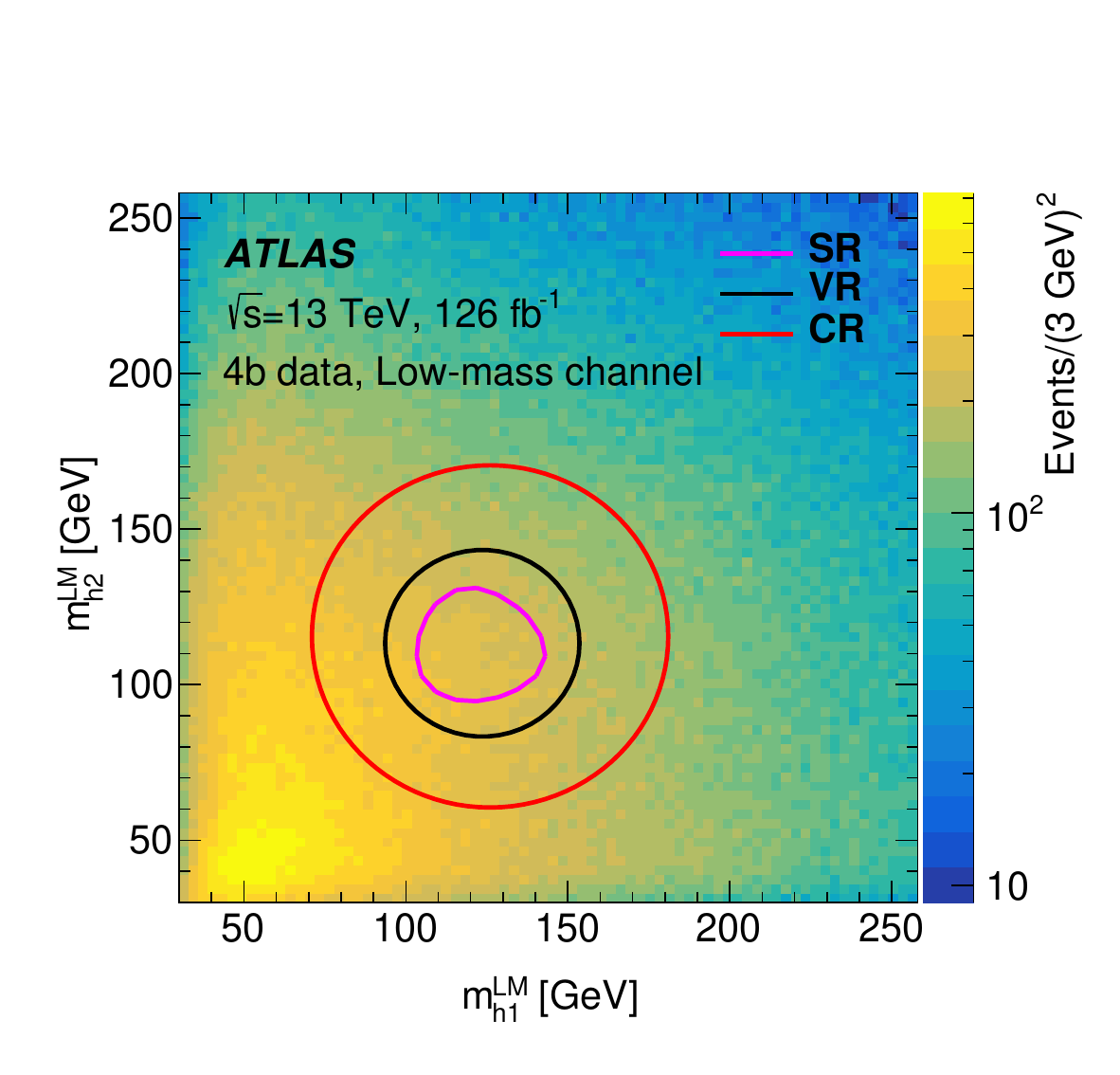}
\label{fig:lm_mass_data}
}~
\subfloat[$m_{\susy{H}}=\SI{250}{\gev}$]{
\centering
\includegraphics[width=0.5\linewidth]{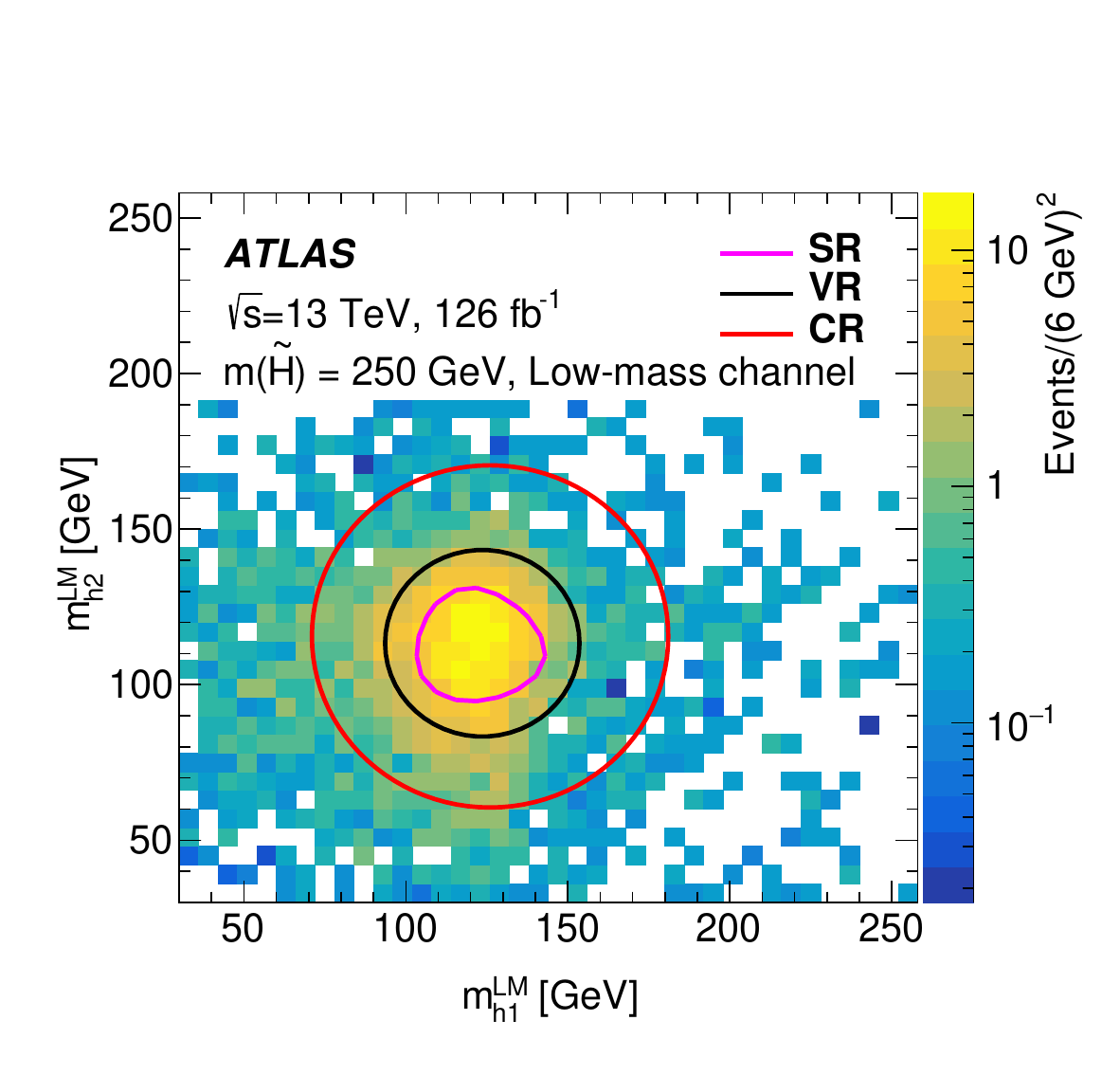}
\label{fig:lm_mass_sig}
}
\caption{Reconstructed Higgs boson masses in the low-mass channel. The left plot shows the $4b$ data, while the right plot shows the $4b$ MC signal sample for a higgsino mass of $\SI{250}{\gev}$. The red, black, and pink (i.e., innermost) contours correspond to the outer boundaries of the control, validation, and signal regions respectively.}
\label{fig:lm_mass}
\end{figure}
 
\subsubsection{Background estimation strategy}
 
The dominant background in the low-$\met$ regime of the low-mass channel comes from QCD multijet processes. Because these processes are difficult to model sufficiently well in simulation, a purely data-driven approach is used to estimate the background from all sources. This estimate makes use of an alternative set of regions with exactly two $b$-jets and two or more non-$b$-tagged jets instead of four or more $b$-jets. All other analysis selections are the same as those previously described in Sec.~\ref{sec:lm_ev}. These alternative regions, called the \enquote{$2b$} sample, have more events than provided by the nominal $4b$ sample while also having significantly less signal contamination.
 
The Higgs boson candidates for the $2b$ sample are reconstructed using the two $b$-jets plus two jets selected randomly from the event's non-$b$-tagged jets. The additional jets are required to have $|\eta| < 2.5$ to match the $\eta$ requirements for $b$-tagging. The mass values of the reconstructed Higgs bosons are used to create the VRs and CRs. The VRs are defined for each data-taking period by $R_{hh}^\text{VR} < \SI{30}{\gev}$ and $X_{hh}^\text{SR} \ge 1.6$, where $R_{hh}^\text{VR}$ is given by
\begin{equation*}
R_{hh}^\text{VR}=\sqrt{(m(h_1^\text{LM})-\SI{123.6}{\gev})^2+(m(h_2^\text{LM})-\SI{113.3}{\gev})^2}.
\end{equation*}
 
The CRs are defined by $R_{hh}^\text{CR} < \SI{55}{\gev}$ and $R_{hh}^\text{VR} \ge \SI{30}{\gev}$, where $R_{hh}^\text{CR}$ is given by
\begin{equation*}
R_{hh}^\text{CR}=\sqrt{(m(h_1^\text{LM})-\SI{126.0}{\gev})^2+(m(h_2^\text{LM})-\SI{115.5}{\gev})^2}.
\end{equation*}
 
The VRs and CRs form ellipses around the SRs with the central values shifted by factors of 1.03 and 1.05, respectively. These shifts avoid over-representing low ($m(h_1^\text{LM}),\,m(h_2^\text{LM})$) values, which have a higher density of events. The SRs, VRs, and CRs are each defined for both the $2b$ and $4b$ samples. The $2b$ CRs, $4b$ CRs, and $2b$ SRs are used to create a background model for the $4b$ SRs, where the search is performed. Figure~\ref{fig:lm_mass} shows the $m(h_1^\text{LM})$ versus $m(h_2^\text{LM})$ distributions for both the $4b$ data and the $\SI{250}{\gev}$ signal simulation with the region definitions overlaid.
 
The background model is created by normalizing and reweighting the $2b$ SR to estimate the behavior of the background in the $4b$ SR. The normalization factor is determined from the CRs as
\begin{equation*}
\mu_\text{CR}=\frac{N_\text{CR}^{4b}}{N_\text{CR}^{2b}},
\end{equation*}
where $N_\text{CR}^{4b}$ ($N_\text{CR}^{2b}$) is the observed number of data events in the $4b$ ($2b$) CR. The parameter $\mu_\text{CR}$ is measured separately for the 2016, 2017, and 2018 data-taking periods because of the differences in the triggers used to record the data.

The $2b$ regions are reweighted using BDTs~\cite{Rogozhnikov:2016bdp} to correct for any kinematic differences between the $2b$ and $4b$ regions. The BDTs are trained on data events in the CRs separately for the 2016, 2017, and 2018 data-taking periods. The hyperparameters used in common for each year are the maximum number of layers (5), the minimum number of events per node (250), the learning rate (0.3), and the sampling fraction (0.4). The number of trees was set to 50 for 2016, 75 for 2017, and 100 for 2018. A greater number of trees are used for later years because the data samples are larger. A large set of 51 input variables is provided, as the BDT training can select which variables to weight most strongly. These include the mass, energy, $\pt$, $\eta$, and $\phi$ of each Higgs boson candidate and Higgs boson candidate jet; the mass and $\pt$ of the di-Higgs system; the number of jets; the $\met$; $X_{W\:\!\!t}$ without the light-jet requirement; the number of track-jets associated with each Higgs boson candidate; and 14 angular variables.
 
For each tree of the BDT, events are separated into two nodes using the input variable and the cut value maximizing the $\chi^2$ difference between the nodes. This process repeats at each resulting node until either the maximum number of layers or minimum number of events per node is reached.
 
Each of the final nodes (or leaves) is given a weight
\begin{equation*}
w = \textrm{e}^{\lambda\cdot\text{log}(N_\text{CR,leaf}^{4b}/N_\text{CR,leaf}^{2b})},
\label{susyeq:bdtweight}
\end{equation*}
where $\lambda$ is the learning rate and $N_\text{CR,leaf}^{4b}$ ($N_\text{CR,leaf}^{2b}$) is the number of events in the leaf for the $4b$ ($2b$) CR. The results are then entered into a new decision tree. This process iterates until the desired number of trees is reached. The final weight for each event is the product of the event's weight in each tree, normalized such that the number of events in the reweighted $2b$ CR matches that of the $4b$ CR. After training on data events in the CR, the BDT is used to calculate weights for all $2b$ events. The reweighted $2b$ SR then forms the background estimate for the $4b$ SR.
 
A bootstrapping procedure is used to increase the stability of the BDT reweighting and estimate its statistical uncertainty. In addition to the nominal BDT, an ensemble of 100 BDTs was trained for each year. These BDTs were trained using the same input events but with random Poisson weights with a mean of 1 to model the effect of statistical variations on the training data. The background estimate in any region of phase space is then set to the value obtained from the median of the $100+1$ variations. Half of the difference between the 84th percentile and 16th percentile of the $100+1$ bootstrap variations is taken to be the statistical uncertainty due to the reweighting. This is combined in quadrature with the Poisson uncertainties from the $2b$ samples.

Figure~\ref{LMfig:VRmeff} shows the $\meff$ distribution in the VR for the 2017 dataset before and after the reweighting is applied. Excellent agreement is observed between the reweighted $\meff$ distributions of the $2b$ and $4b$ data. Good agreement was also found in all cases for other years and input variables.

\begin{figure}
\subfloat[Before reweighting]{
\centering
\includegraphics[width=0.5\linewidth]{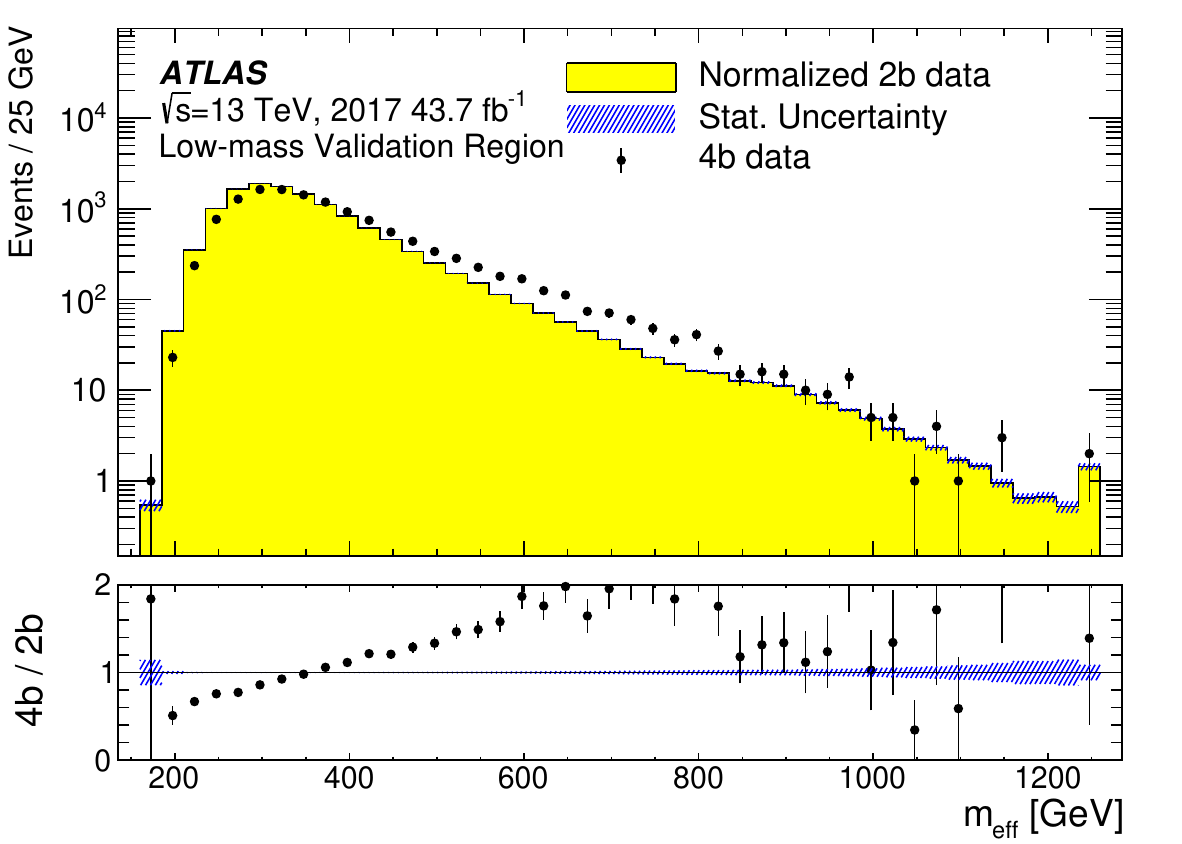}
\label{LMfig:VRmeff_1}
}~
\subfloat[After reweighting]{
\centering
\includegraphics[width=0.5\linewidth]{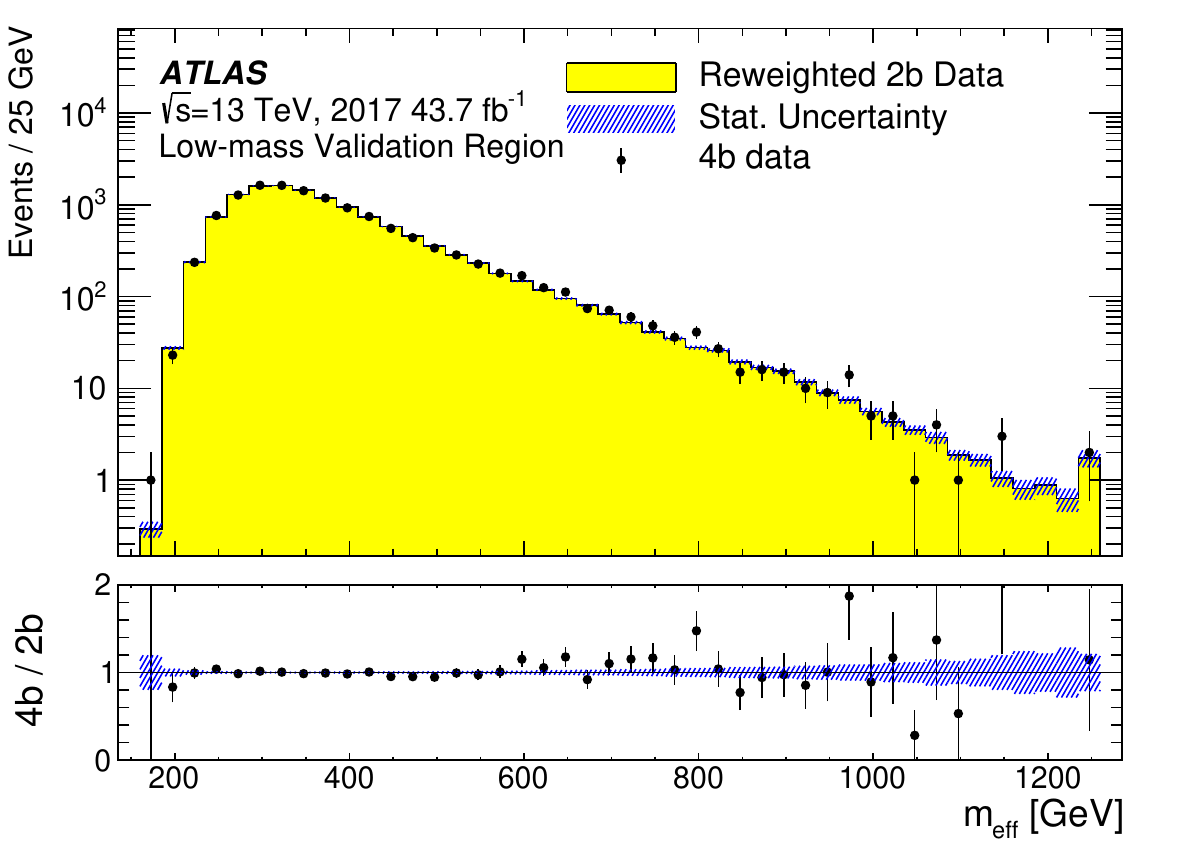}
\label{LMfig:VRmeff_2}
}

\caption{Comparison of the $2b$ and $4b$ \meff distributions in the 2017 VR of the low-mass channel. The left plot shows the \meff distribution before BDT reweighting. The right plot illustrates the same distribution, with the $2b$ events having the reweighting applied to match the $4b$ data. The final nonempty bin contains overflow events. The $2b$ distribution is normalized to the $4b$ integral to show the shape differences. The blue shading shows the uncertainty due to the $2b$ sample size for the left plot and both the reweighting and the $2b$ sample size for the right plot.}
\label{LMfig:VRmeff}
\end{figure}



\section{Systematic uncertainties}
\label{sec:systematics}

Systematic uncertainties from various sources are evaluated for this analysis. Uncertainties are divided into three types: experimental uncertainties, theoretical uncertainties, and uncertainties in the data-driven background estimate.  Experimental uncertainties quantify systematic effects due to the ATLAS detector and LHC conditions. Theoretical uncertainties account for possible mismodeling in the simulations of various physical processes. Experimental and theoretical uncertainties apply to the signal models used by the two channels as well as to the MC background estimates for the high-mass channel. Uncertainties in the data-driven background estimate are applied to the low-mass channel to account for potential mismodeling from the normalization factors and the BDTs used to estimate the background.
 
\subsection{Experimental uncertainties}
Systematic uncertainties associated with the jet energy scale and resolution~\cite{JETM-2018-05}, jet mass scale~\cite{JETM-2018-02}, flavor-tagging efficiencies~\cite{FTAG-2018-01}, trigger efficiencies, soft $\met$ terms~\cite{PERF-2016-07}, \pileup conditions~\cite{PERF-2014-03, STDM-2015-05}, and luminosity are included in this analysis. Apart from the 1.7\% uncertainty in the combined 2015--2018 integrated luminosity~\cite{ATLAS-CONF-2019-021}, obtained using the LUCID-2 detector~\cite{LUCID2} for the primary luminosity measurements, these uncertainties are assessed separately for each signal and background process. For the high-mass channel, experimental uncertainties are correlated across all of the SRs, VRs, and CRs for the higgsino mass hypothesis being tested. For the low-mass channel, these uncertainties are only applied to the $4b$ SRs as no other regions are used in the fit. For the high-mass channel, the jet energy scale, jet energy resolution, and jet mass scale uncertainties are estimated using a single enlarged CR, VR, and SR for each mass point. These three regions are created for each mass point by merging each of the mass point's CRs, VRs, and SRs, respectively, and are used to avoid large statistical fluctuations that could cause instabilities in the fit. For the low-mass channel, the trigger and offline flavor-tagging uncertainties are calculated jointly in order to properly account for correlations. The jet energy resolution uncertainties for the low-mass channel are smoothed with a Gaussian kernel to mitigate the effect of statistical fluctuations.
 
\subsection{Theoretical uncertainties}
\label{sec:theory_syst}
The signal models used by the two channels are also subject to theoretical uncertainties. MC generator-level samples for a representative set of mass points are used to assess the effect of varying the factorization and renormalization scales, merging scale, parton shower tuning, and initial- and final-state radiation parameters. The differences resulting from these variations are applied as the systematic uncertainties. For the low-mass channel, these differences are smoothed using a Gaussian kernel.
 
Theoretical uncertainties are also applied to the background samples for the high-mass channel. Uncertainties quantifying the systematic effects due to possible mismodeling associated with the MC generator and parton shower models, as well as with the modeling of initial- and final-state radiation, are applied to the \ttbar and single-top backgrounds. For the $tW$ process, the uncertainty associated with interference with \ttbar production was estimated by comparing the nominal sample with an alternative sample generated using the diagram subtraction scheme~\cite{Frixione:2008yi,ATL-PHYS-PUB-2016-020}. Uncertainties related to the renormalization, factorization, resummation, and matching scales are applied to the ${W}+$jets and ${Z}+$jets backgrounds. Uncertainties of 50\% are applied to the $\ttbar$+$X$ and diboson backgrounds. Since the $\ttbar$+$X$ and diboson backgrounds make up a small fraction of the overall background, these uncertainties have no significant impact on the analysis.
 
Figure~\ref{fig:hm_systs} illustrates the total postfit uncertainties in the high-mass channel's SRs, as well as the dominant components in each. Statistical uncertainties dominate. At masses below $\SI{1000}{\gev}$, theoretical uncertainties associated with the generation and parton showering of \ttbar events are one of the most important systematic uncertainty components, due to the dominance of this background process. In contrast, signal regions with $m_{\susy{H}} \ge \SI{1000}{\gev}$ are more affected by uncertainties in $Z$+jets processes because of the increasing importance of the $Z\rightarrow \nu\nu$ background at higher \met.
 
\begin{figure}[h!]
\centering
\includegraphics[width=0.8\textwidth]{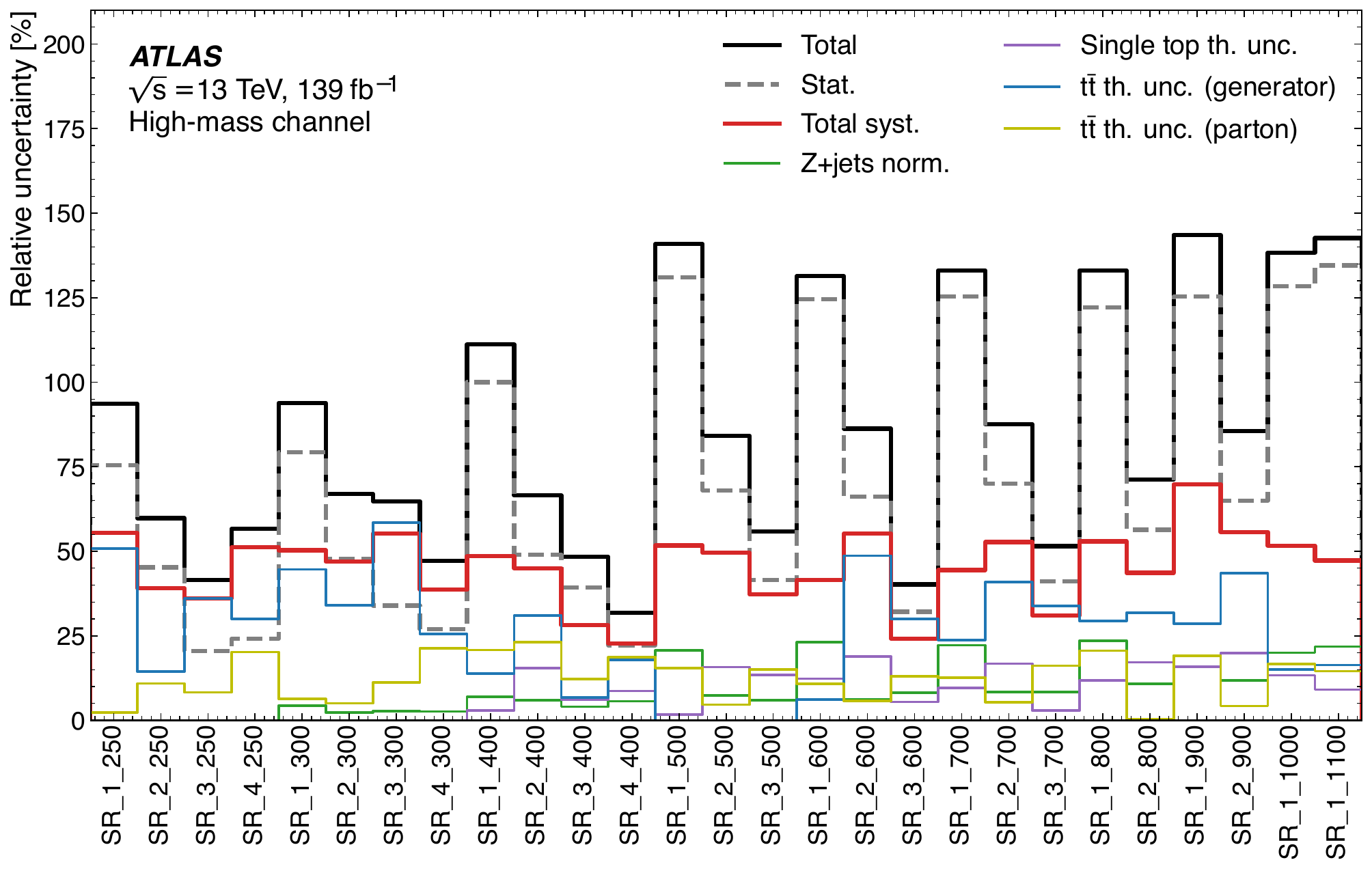}
\caption{Breakdown of the dominant postfit systematic uncertainties in the SRs of the high-mass channel. Because of their correlations, the total systematic uncertainty may exceed the quadrature sum of its components. The total uncertainty shown is the quadrature sum of the statistical and total systematic uncertainties.}
\label{fig:hm_systs}
\end{figure}
 
\subsection{Data-driven background uncertainties}
In the low-mass channel, three sources of systematic uncertainty affect the data-driven background estimate. Their effects are estimated separately for the 2016, 2017, and 2018 data-taking periods.
 
A nonclosure uncertainty is assigned to account for imperfections in the reweighting procedure. This is estimated by evaluating the bin-by-bin differences between the $4b$ and reweighted $2b$ samples in the CRs. A nuisance parameter corresponding to the discrepancy is assigned to each bin where the difference is greater than the combined statistical uncertainty of the $2b$ and $4b$ samples. No nuisance parameters are assigned to bins where the discrepancy is less than this uncertainty. The nonclosure uncertainties are treated as uncorrelated in the statistical analysis.
 
A transfer uncertainty is assigned to the extrapolation of results from the CRs to the SRs. This transfer uncertainty is divided into a normalization component and a shape component. To estimate the transfer shape uncertainty, the BDTs were retrained using data from the VRs instead of the CRs. All other aspects of the training, including the bootstrapping procedure, were unchanged. This creates alternative background estimates with weights derived closer to the SRs. Each VR-derived estimate is then normalized to the corresponding nominal (CR-derived) estimate in the SR to separate this uncertainty from the normalization uncertainty. The difference between the two estimates is calculated for each bin and applied as an uncertainty if the difference is greater than the combined statistical uncertainty of the estimates. As with the nonclosure uncertainty, the transfer shape uncertainty is treated as uncorrelated across bins and is not applied to bins where the difference is less than the combined statistical uncertainty of the estimates.
 
The third source is the transfer normalization uncertainty, which is computed using two methods. In the first method, the uncertainty is taken to be the fractional difference between the numbers of SR events predicted by using the nominal and VR-derived background estimates. In the second method, the uncertainty for each data-taking period is taken to be the fractional difference between the predicted and observed numbers of events in the VR. The normalization uncertainty is taken to be the larger of its two estimated values.
 
A slightly different treatment is used to assign transfer uncertainties to the discovery regions. Since the discovery regions each consist of a single bin, the transfer normalization and shape uncertainties are calculated jointly as the difference between the numbers of SR events predicted by using the nominal and VR-derived background estimates. In addition, the VR nonclosure is calculated as the difference between the numbers of predicted and observed events in the VR. The transfer uncertainties for the discovery regions are then taken to be the larger of the calculated transfer and VR nonclosure uncertainties. The bootstrap and nonclosure uncertainties are assigned in the same way as in the exclusion regions, except that the nonclosure uncertainty is applied even if it is less than the statistical uncertainty.
 
The components of the background uncertainty for the low-mass channel are shown in Fig.~\ref{LMfig:bkgsyst}. Considering all sources of systematic uncertainty, the nonclosure, jet energy resolution, and jet flavor composition uncertainties tend to have the greatest effect,
the last being a component of the jet energy scale uncertainty.
 
A 50\% uncertainty is applied to the QCD multijet prediction for the high-mass channel to account for potential mismodeling from the NN-assisted reweighting.
 
\begin{figure}[h!]
\includegraphics[width=0.95\linewidth]{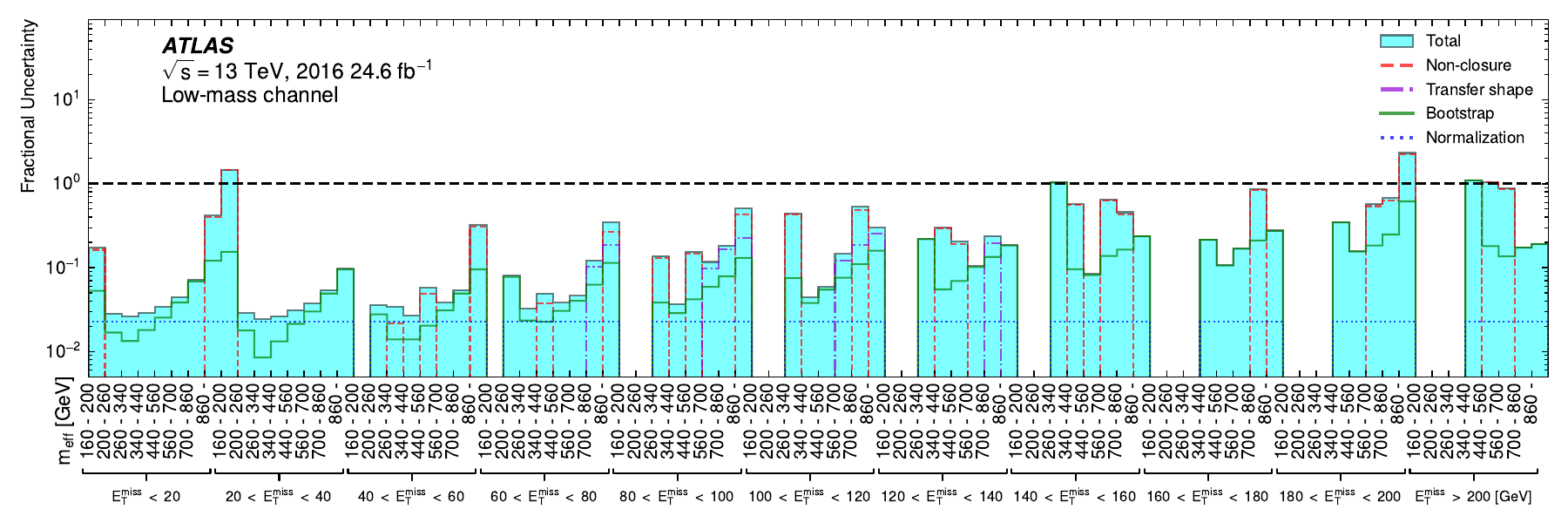}
\includegraphics[width=0.95\linewidth]{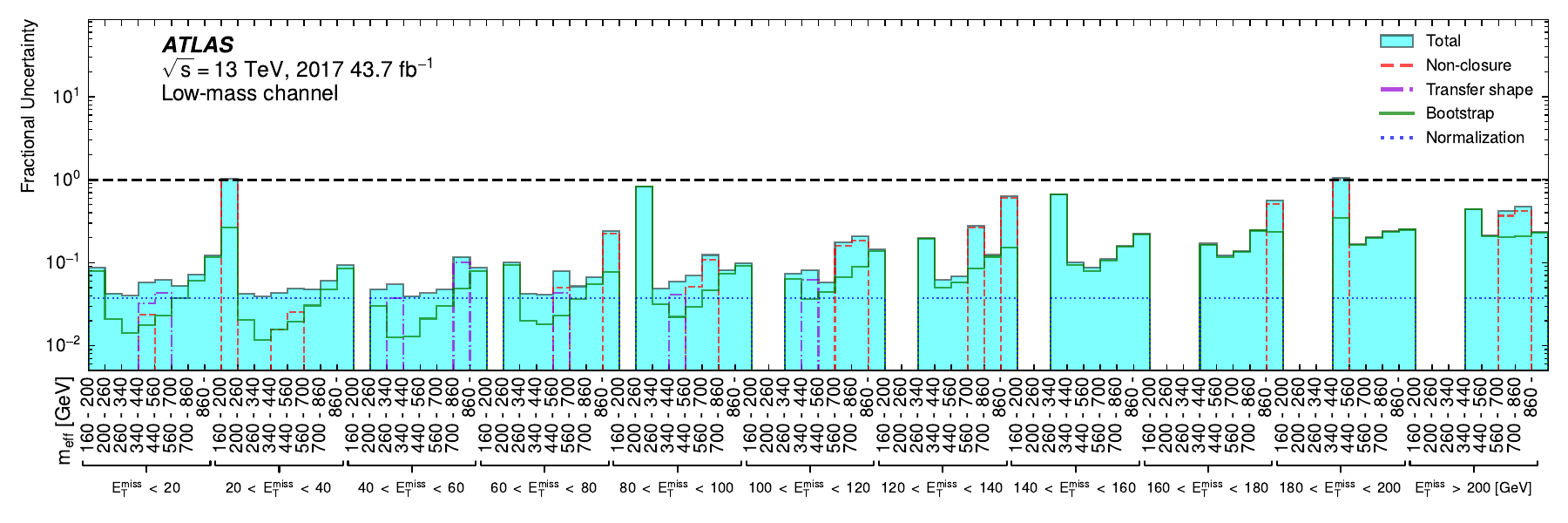}
\includegraphics[width=0.95\linewidth]{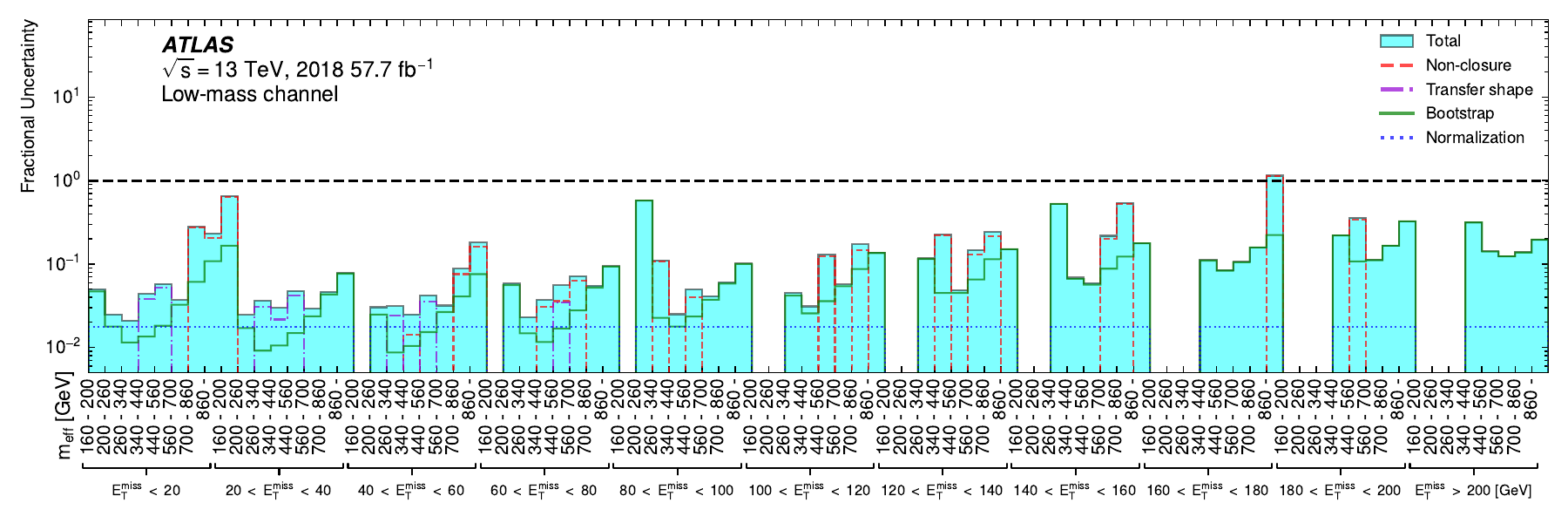}
 
\caption{Background uncertainties as a fraction of background yields for the 2016, 2017, and 2018 SRs. Prefit values are shown. The bootstrap component also includes the Poisson component of the statistical uncertainty of the $2b$ sample. No uncertainties are shown for bins predicted to have no background events. The total uncertainty shown is the quadrature sum of the individual uncertainties. All uncertainties are treated as symmetric.}
\label{LMfig:bkgsyst}
\end{figure}


\section{Results}
\label{sec:result}

 
\subsection{High-mass channel}
The level of agreement between the data and the postfit background prediction is shown for each VR of the high-mass channel in Fig.~\ref{HMfig:final_vr}. All data observations are found to be in agreement with the background prediction within the analysis uncertainties. The \ttbar and $Z$+jets normalization factors are measured to be between 0.9 and 1.3 for all mass hypotheses.
\begin{figure}
\centering
\includegraphics[width=\textwidth]{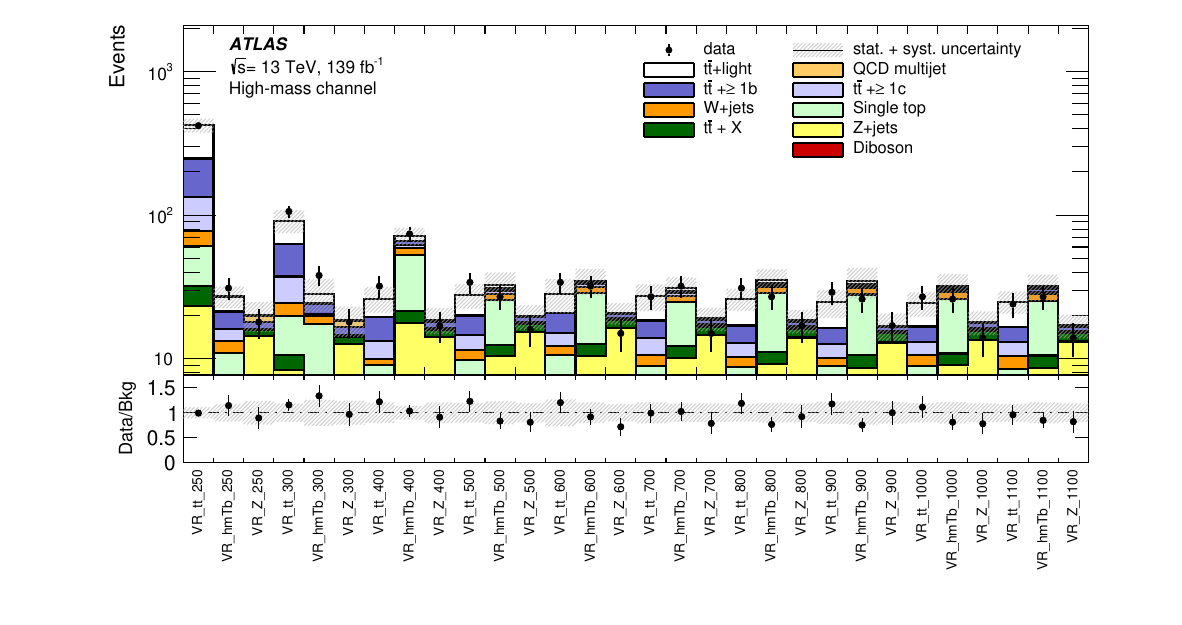}
\caption{Postfit VR yields in the high-mass channel. The upper panel shows the observed number of events, as well as the postfit background predictions in each region. The bottom panel shows the ratio of the observed data to the total background prediction. The shaded areas correspond to the total statistical and systematic uncertainties obtained after the fit and described in Sec.~\ref{sec:systematics}.}
\label{HMfig:final_vr}
\end{figure}

\begin{figure}
\centering
\includegraphics[width=\textwidth]{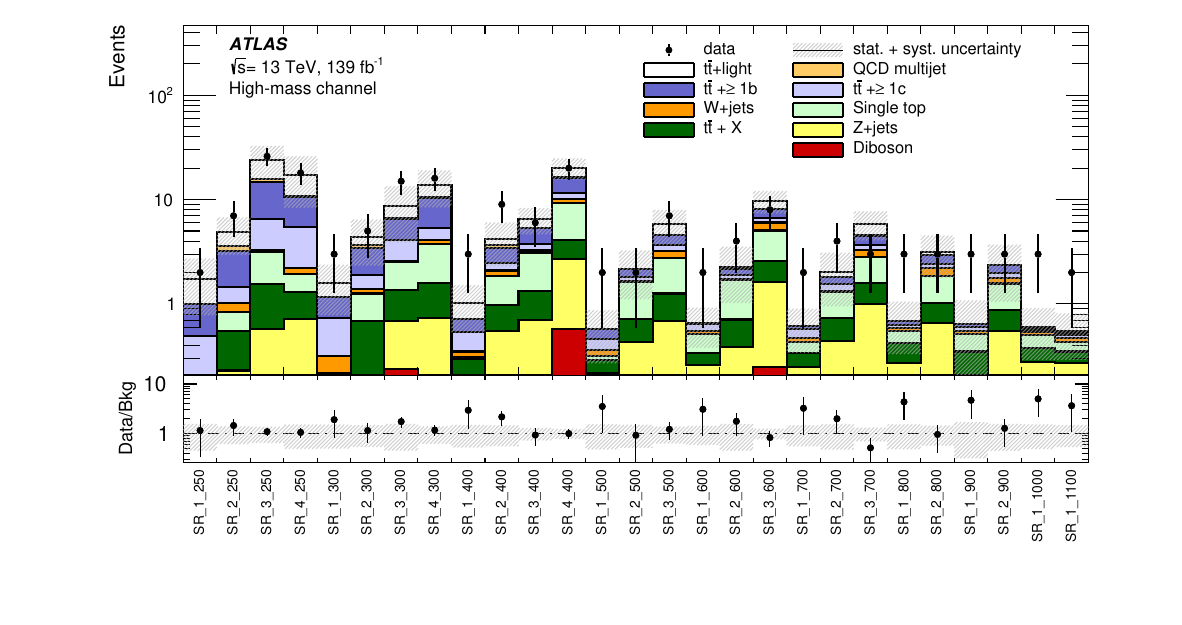}
\caption{Postfit SR yields in the high-mass channel. The upper panel shows the observed number of events, as well as the postfit background predictions in each region. The bottom panel shows the ratio of the observed data to the total background prediction. The shaded areas correspond to the total statistical and systematic uncertainties obtained after the fit and described in Sec.~\ref{sec:systematics}.}
\label{HMfig:final_sr}
\end{figure}
 
The SR observations are shown in Fig.~\ref{HMfig:final_sr}, together with the postfit background predictions. No statistically significant excess is found, and the largest excess is observed in \texttt{SR\_1\_1000} with a local statistical significance of $1.9\sigma$. The excesses seen in \texttt{SR\_1\_900}, \texttt{SR\_1\_1000} and \texttt{SR\_1\_1100} are, however, similar due to the highly correlated SR definitions. The same three data events are observed in the first two of these SRs, and one of those events fails the \texttt{SR\_1\_1100} selection.

\subsection{Low-mass channel}
 
The observations and normalized background predictions are shown for each $\met$ and $\meff$ bin of the VRs in Fig.~\ref{LMfig:unfolded_vr}. The lower panel shows the significance of the differences, using the profile likelihood method from Ref.~\cite{Cousins:2007bmb}. No significant mismodeling is observed.
 
\begin{figure}[ph]
\includegraphics[width=0.92\linewidth]{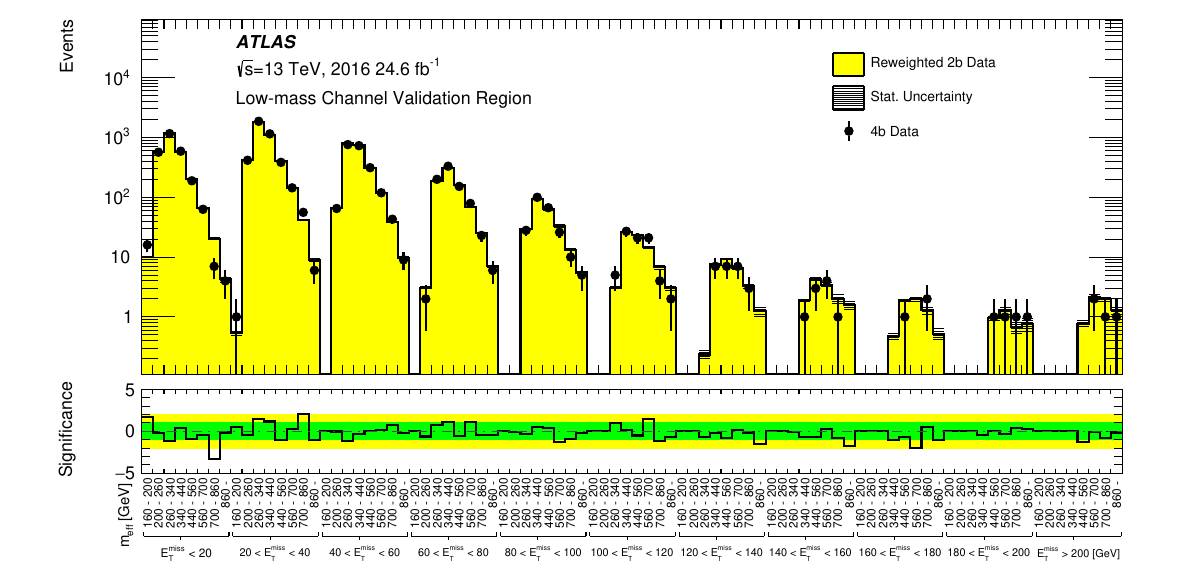}
\includegraphics[width=0.92\linewidth]{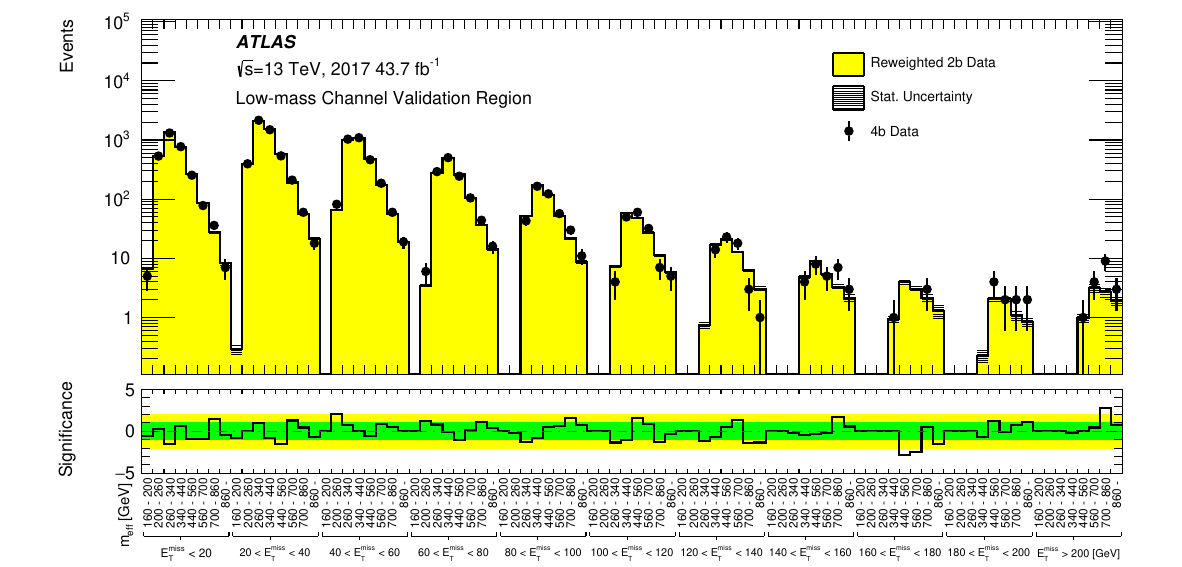}
\includegraphics[width=0.92\linewidth]{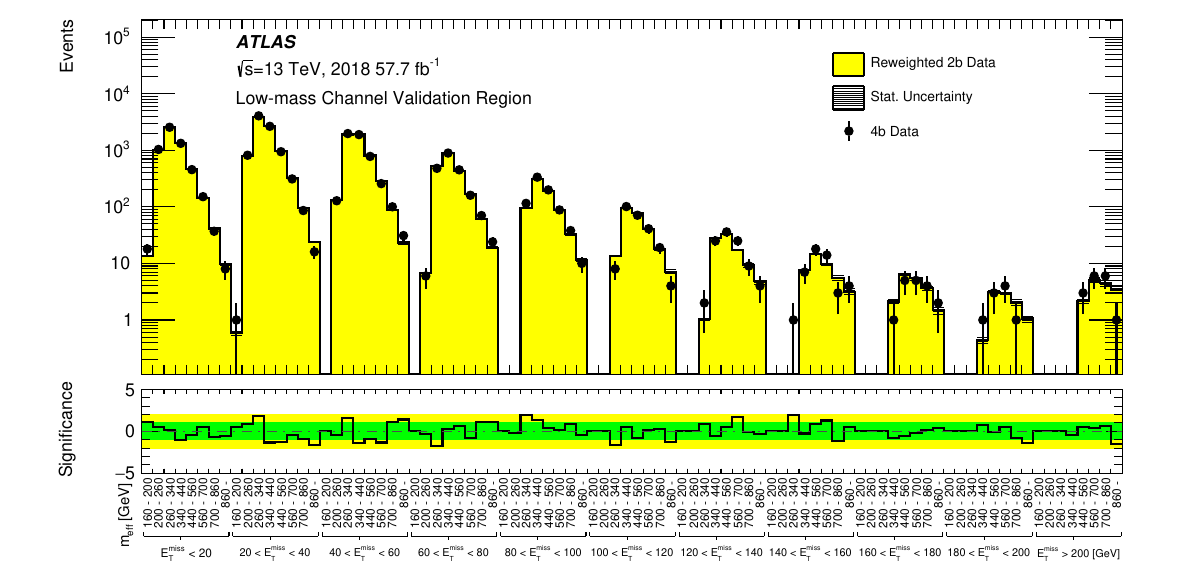}
 
\caption{Data and background (reweighted $2b$) predictions for each $4b$ VR $\met$ and $\meff$ bin of the low-mass channel for the 2016, 2017, and 2018 data-taking periods. The background is normalized to the $4b$ data. The bottom panel shows the significance of any differences between the observed $4b$ data and the background prediction. The $1\sigma$ and $2\sigma$ bands are shown in green and yellow, respectively.}
\label{LMfig:unfolded_vr}
\end{figure}
 
Figure~\ref{LMfig:unblinded_sr} shows the results for each $\met$ and $\meff$ bin of the SRs. No significant excess above the Standard Model prediction is observed. The largest upward fluctuation in the 213 kinematically allowed bins is observed in the $\met > \SI{200}{\gev}$, $\meff > \SI{860}{\gev}$ bin for the 2017 data-taking period. This bin has six observed events and $1.51 \pm 0.35$ predicted background events, corresponding to a local significance of $2.6\sigma$. This excess occurs for events with high $\met$, where the high-mass channel has greater sensitivity.
 
\begin{figure}[ph]
\includegraphics[width=0.92\linewidth]{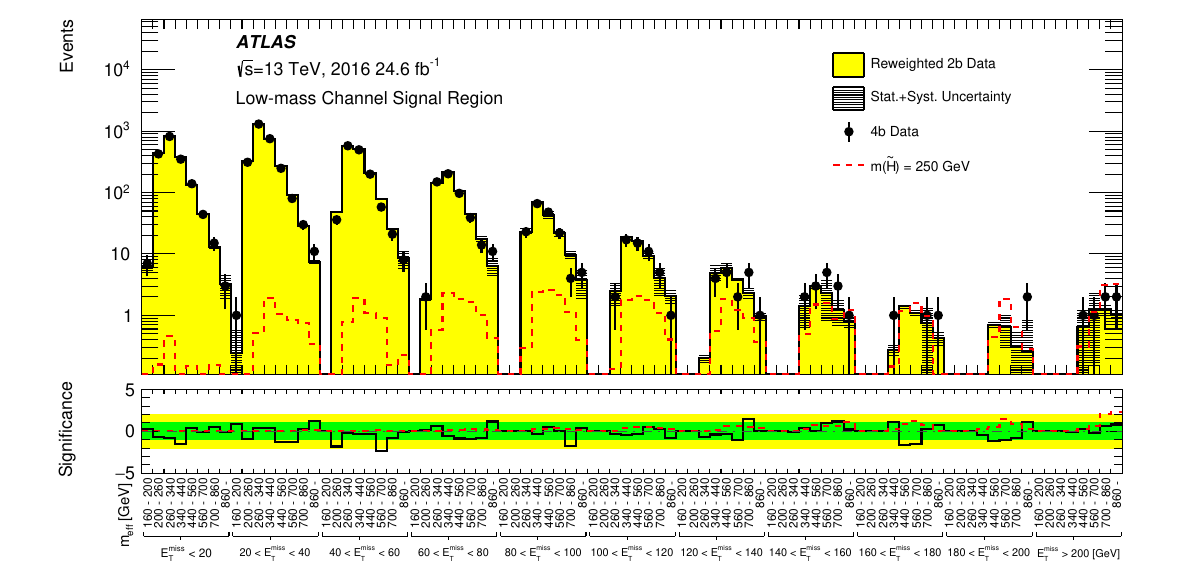}
\includegraphics[width=0.92\linewidth]{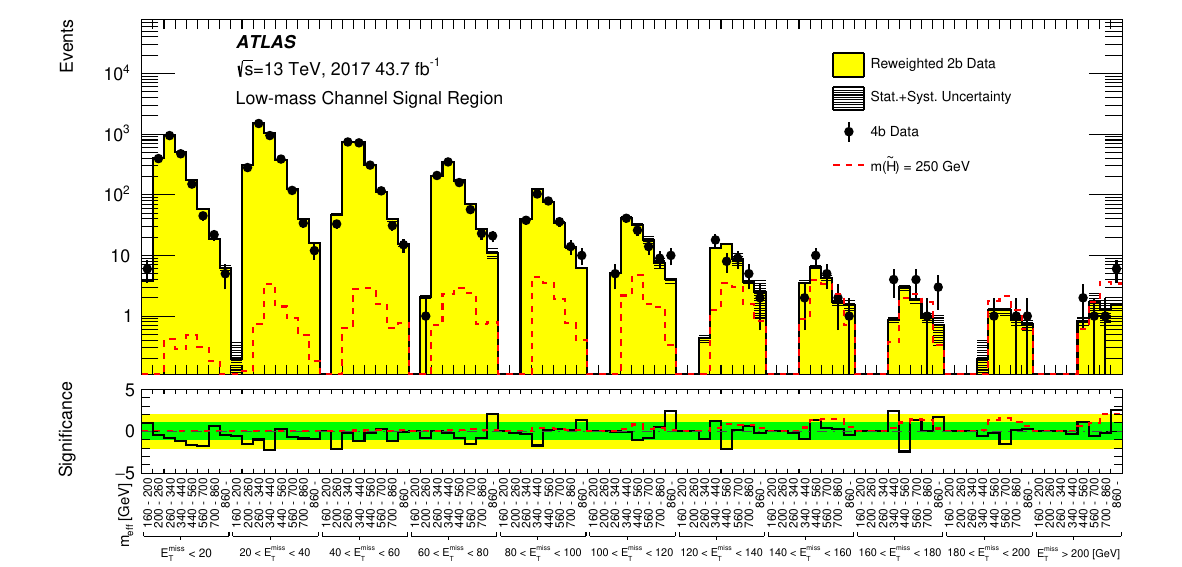}
\includegraphics[width=0.92\linewidth]{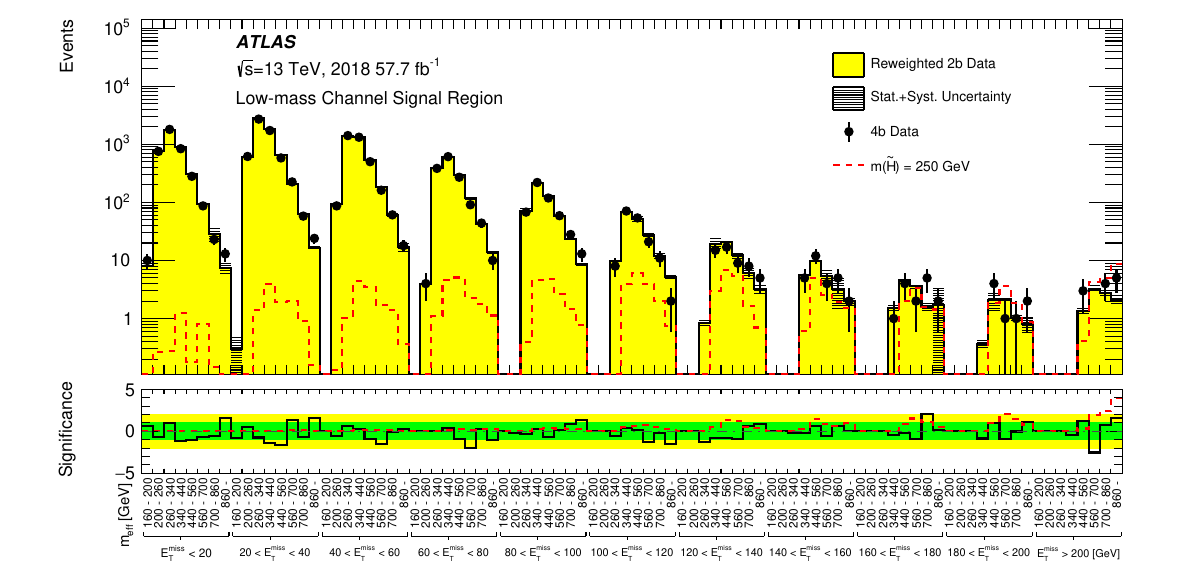}
\caption{Prefit data and background (reweighted $2b$) predictions for each $4b$ SR $\met$ and $\meff$ bin of the low-mass channel for the 2016, 2017, and 2018 data-taking periods. The bottom panel shows the significance of any differences between the observed $4b$ data and the background prediction. The $1\sigma$ and $2\sigma$ bands are shown in green and yellow, respectively.}
\label{LMfig:unblinded_sr}
\end{figure}


\section{Statistical interpretation}
\label{sec:stats}

Following the Neyman--Pearson lemma, upper limits on higgsino pair-production cross sections are set using test statistics~\cite{Cowan:2010js} based on the profile likelihood ratio. The $p$-values of the statistical tests are obtained by following the \CLs prescription~\cite{Read:2002hq} and, unless stated otherwise, using the asymptotic approximation described in Ref.~\cite{Cowan:2010js}. Results obtained using the asymptotic approximation were confirmed with pseudo-experiments. The HistFitter framework~\cite{Baak:2014wma} is used for the high-mass channel and the pyhf framework~\cite{pyhf, pyhf_joss} is used for the low-mass channel. Two different types of upper limits are provided for each channel: model-independent limits and model-dependent limits for the various simulated higgsino masses $m_{\susy{H}}$.
 
\subsection{Model-independent limits}
Model-independent limits on the number of beyond-the-SM physics events are set for each discovery SR of the analysis. The upper limits are obtained by assuming no signal contamination in the analysis CRs, and assigning to each SR a number of signal events corresponding to the fit's BSM signal strength $\mu$. The different backgrounds and model-independent signals are then adjusted to determine the 95\% confidence level (CL) upper limits. Results are calculated using 5000 pseudoexperiments for the high-mass discovery regions \texttt{SR\_1\_250}, \texttt{SR\_1\_500}, and \texttt{SR\_1\_1000} and 20\;\!000 pseudoexperiments for the low-mass discovery regions \texttt{SR\_LM\_150} and \texttt{SR\_LM\_300}.
The upper limits for the discovery regions of the two channels are summarized in Table~\ref{tab:HM_ModelIndpLimits}.
 
\begin{table}
\centering
\caption{
Results from the high-mass and low-mass channel discovery regions. The columns include, from left to right: the observed number of events $N_\text{obs}$, predicted number of events $N_\text{pred}$, and 95\% CL upper limits on the visible cross section
($\langle\epsilon\sigma\rangle_{\textrm{obs}}^{95}$) and on the number of
signal events ($S_{\textrm{obs}}^{95}$ ). The $S_{\textrm{exp}}^{95}$ column
shows the 95\% CL upper limit on the number of
signal events, given the expected number (and $\pm 1\sigma$
excursions of the expectation) of background events.
The last column
indicates the discovery $p$-value (denoted by $p(s = 0)$) and its corresponding significance in parentheses. The $p$-values are capped at 0.5.}
\label{tab:HM_ModelIndpLimits}
\begin{tabular}{lcccccc}
\hline{Signal channel} & $N_\text{obs}$ & $N_\text{pred}$                       & $\langle\epsilon{\sigma}\rangle_{\textrm{obs}}^{95}$\,[fb]  &  $S_{\textrm{obs}}^{95}$  & $S_{\textrm{exp}}^{95}$ & $p(s=0)$  \\
\hline
\texttt{SR\_1\_250}  & \num{2} & \num{1.8\pm 1.0} & $0.04$ &  $6.2$ & $ { 5.9 }^{ +1.7 }_{ -0.9 }$ & $ 0.48$~$(0.05)$ \\%
\texttt{SR\_1\_500}  & \num{2} & \num{0.58\pm 0.30}  & $0.04$ &  $5.5$ & $ { 4.0 }^{ +1.7 }_{ -0.6 }$ & $ 0.18$~$(0.92)$ \\%
\texttt{SR\_1\_1000}  & \num{3} & \num{0.60\pm 0.31}  & $0.05$ &  $6.7$ & $ { 4.3 }^{ +0.9 }_{ -0.9 }$ & $ 0.03$~$(1.9)$\phantom{0} \\%
\hline
\texttt{SR\_LM\_150} & \num{1790}\phantom{000} & \num{1860\pm 50}\phantom{00} & $0.73$ & $92$\phantom{0.0} & \phantom{.}$ { 127 }^{ +48 }_{ -34 }$\phantom{$0_{.}$} & $ 0.5$~\phantom{0}$(0.00)$ \\
\texttt{SR\_LM\_300} & \num{97}\phantom{0} & \num{77.0\pm 5.3}\phantom{0} & $0.31$ & $39$\phantom{0.0} & \phantom{.}$ { 22 }^{ +9 }_{ -6 }$\phantom{$_{.0}$} & $ 0.03$~$(1.8)$\phantom{0} \\
\bottomrule
\end{tabular}
\end{table}
 
\begin{figure}
\centering
\subfloat[]{
\includegraphics[width=0.5\textwidth]{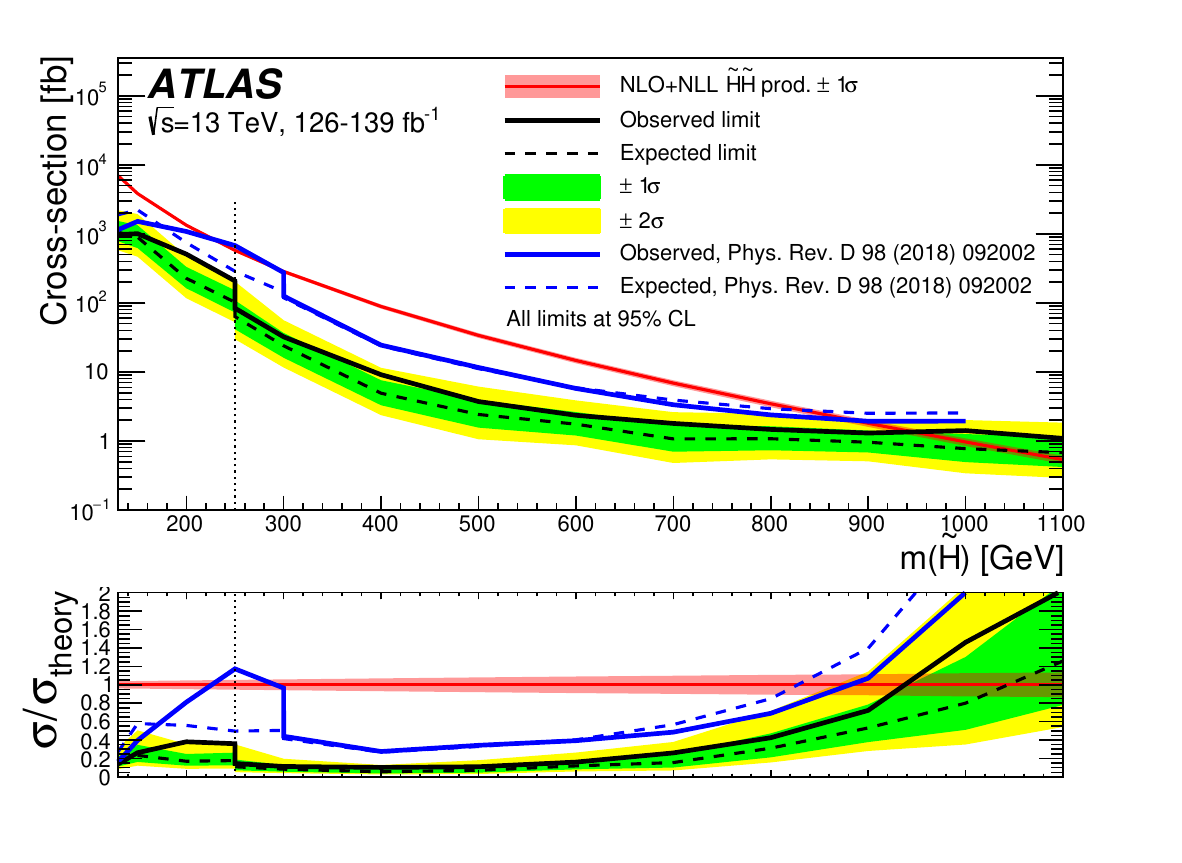}
\label{fig:limits_a}
}~
\subfloat[]{
\includegraphics[width=0.5\textwidth]{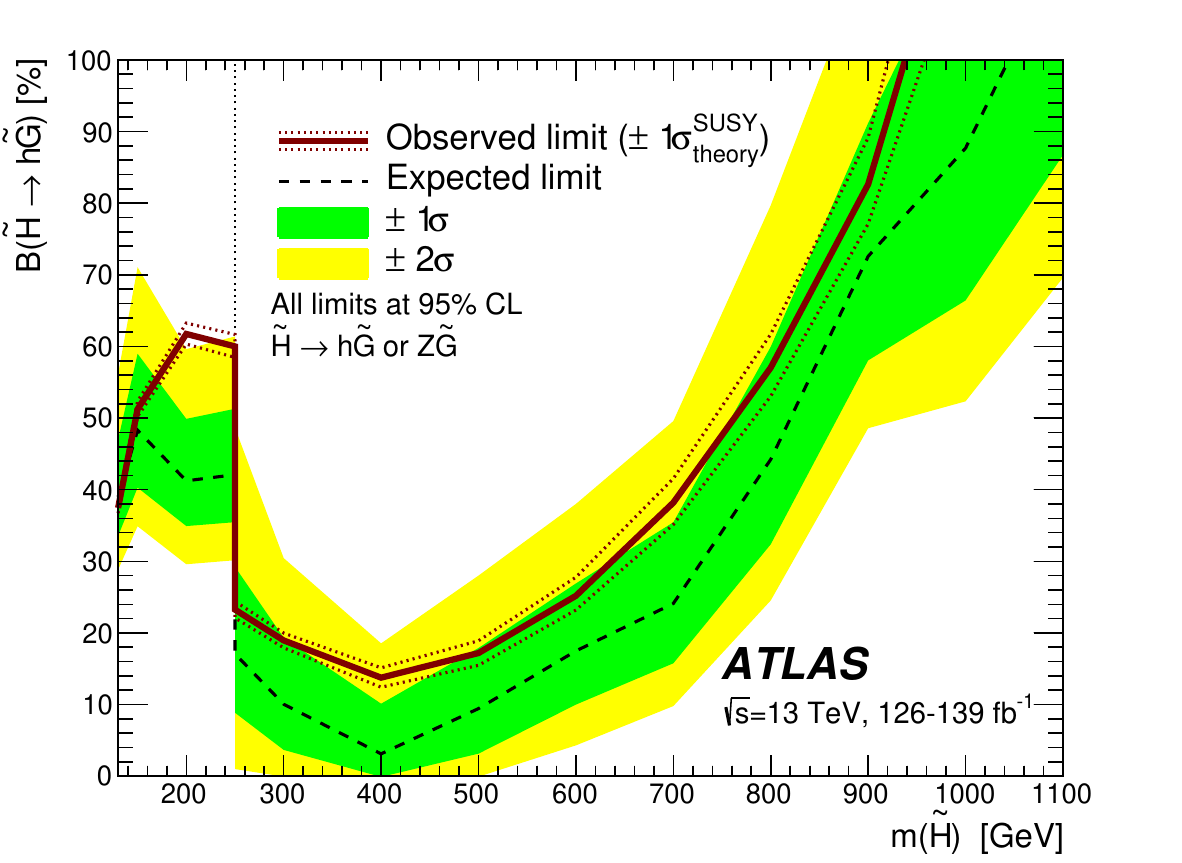}
\label{fig:limits_b}
}
\caption{Exclusion limits in the low-mass and high-mass channels. The low-mass channel is used for $m_{\susy{H}}<\SI{250}{\gev}$, while the high-mass channel is used for $m_{\susy{H}}\ge\SI{250}{\gev}$. The left plot shows the observed (solid) and expected (dashed) 95\% CL upper limits on the higgsino pair-production cross section, assuming a higgsino decay branching ratio of $\mathcal{B}(\susy{H}\rightarrow h + \gravino)=100\%$. The theory cross section and its uncertainty are shown by the solid red line and red shading. Results from a previous ATLAS search using 24.3--36.1\,fb$^{-1}$ of \RunTwo data~\cite{SUSY-2017-02} are shown by the solid (observed) and dashed (expected) blue lines. The bottom panel shows the ratio of the limits to the theory cross section. The right plot shows the 95\% CL observed (solid) and expected (dashed) upper limits on $\mathcal{B}(\susy{H}\rightarrow h + \gravino)$, assuming the theory cross section for higgsino pair production. The higgsinos are assumed to decay as $\susy{H}\rightarrow h + \gravino$ or $\susy{H}\rightarrow Z + \gravino$. For both plots, the phase space above the lines is excluded.}
\label{fig:limits}
\end{figure}

\subsection{Model-dependent exclusion limits}
Since no significant excess above the SM prediction is observed, 95\% CL upper limits are calculated for the simplified signal models described in Sec.~\ref{sec:intro}. Each mass point uses the channel with the better expected limit, with the transition between the low-mass and high-mass channel interpretations occurring at $m_{\susy{H}}=\SI{250}{\gev}$. Figure~\ref{fig:limits_a} shows the upper limit on the higgsino pair-production cross section for a branching ratio of $\mathcal{B}(\susy{H}\rightarrow h + \gravino)=100\%$. Higgsino masses below $\SI{940}{\gev}$ and, owing to the mass of the Higgs boson, above $\SI{130}{\gev}$ are excluded.
 
The results are also interpreted for the case where $\mathcal{B}(\susy{H}\rightarrow h + \gravino)$ is allowed to vary, with all other decays assumed to create $Z$ bosons through $\susy{H}\rightarrow Z + \gravino$. The higgsino pair-production cross section is assumed to match the theoretical prediction. Limits for this interpretation are shown in Fig.~\ref{fig:limits_b}. For the low-mass channel, the $Z$ decays peak outside of the SRs in the $(m(h_1^\text{LM}),m(h_2^\text{LM}))$ plane and would be included in the data-driven background estimate. The limit on the branching ratio is therefore equal to the square root of the limit on $\sigma/\sigma_\text{theory}$ shown in Fig.~\ref{fig:limits_a}. For the high-mass channel, both the Higgs boson and the $Z$ boson decays are included in the signal model, leading to additional sensitivity in the branching ratio plane and the exclusion of $\susy{H}\rightarrow h + \gravino$ branching ratios as low as 14\%.


\section{Conclusion}
\label{sec:conclusion}

A search for pair-produced and mass-degenerate higgsinos decaying to gravitinos and Higgs bosons was performed using the ${\ge}3\,b$-jets\,+\,\met final state. The search uses two separate channels optimized for either low-mass or high-mass higgsinos and exploits the 139~fb$^{-1}$ of $\sqrt{s}=\SI{13}{\tev}$ proton--proton collision data collected by the ATLAS detector during \RunTwo of the LHC. This analysis improves upon the previous ATLAS search in several ways, including using a boosted decision tree to provide enhanced background rejection for the high-mass channel, reoptimizing the $b$-tagging efficiency working point and binning structure for the low-mass channel, implementing a new method for reconstructing Higgs boson candidates, and using improved jet reconstruction and $b$-tagging algorithms.
 
No significant excess above the SM prediction is observed. Higgsinos with masses between $\SI{130}{\gev}$ and $\SI{940}{\gev}$ are excluded at 95\% confidence level for the $\mathcal{B}(\susy{H}\rightarrow h + \gravino)=100\%$ hypothesis. For this hypothesis, this search is the most sensitive to date across the mass range investigated, and provides the most stringent constraints in the 130--800 GeV higgsino mass range. Upper limits are set on $\mathcal{B}(\susy{H}\rightarrow h + \gravino)$, reaching as low as 14\% for a higgsino mass of $\SI{400}{\gev}$. Model-independent limits are also set on the visible cross section for new physics processes.


\section*{Acknowledgements}


We thank CERN for the very successful operation of the LHC and its injectors, as well as the support staff at
CERN and at our institutions worldwide without whom ATLAS could not be operated efficiently.
 
The crucial computing support from all WLCG partners is acknowledged gratefully, in particular from CERN, the ATLAS Tier-1 facilities at TRIUMF/SFU (Canada), NDGF (Denmark, Norway, Sweden), CC-IN2P3 (France), KIT/GridKA (Germany), INFN-CNAF (Italy), NL-T1 (Netherlands), PIC (Spain), RAL (UK) and BNL (USA), the Tier-2 facilities worldwide and large non-WLCG resource providers. Major contributors of computing resources are listed in Ref.~\cite{ATL-SOFT-PUB-2023-001}.
 
We gratefully acknowledge the support of ANPCyT, Argentina; YerPhI, Armenia; ARC, Australia; BMWFW and FWF, Austria; ANAS, Azerbaijan; CNPq and FAPESP, Brazil; NSERC, NRC and CFI, Canada; CERN; ANID, Chile; CAS, MOST and NSFC, China; Minciencias, Colombia; MEYS CR, Czech Republic; DNRF and DNSRC, Denmark; IN2P3-CNRS and CEA-DRF/IRFU, France; SRNSFG, Georgia; BMBF, HGF and MPG, Germany; GSRI, Greece; RGC and Hong Kong SAR, China; ISF and Benoziyo Center, Israel; INFN, Italy; MEXT and JSPS, Japan; CNRST, Morocco; NWO, Netherlands; RCN, Norway; MEiN, Poland; FCT, Portugal; MNE/IFA, Romania; MESTD, Serbia; MSSR, Slovakia; ARRS and MIZ\v{S}, Slovenia; DSI/NRF, South Africa; MICINN, Spain; SRC and Wallenberg Foundation, Sweden; SERI, SNSF and Cantons of Bern and Geneva, Switzerland; MOST, Taipei; TENMAK, T\"urkiye; STFC, United Kingdom; DOE and NSF, United States of America.
 
Individual groups and members have received support from BCKDF, CANARIE, CRC and DRAC, Canada; CERN-CZ, PRIMUS 21/SCI/017 and UNCE SCI/013, Czech Republic; COST, ERC, ERDF, Horizon 2020, ICSC-NextGenerationEU and Marie Sk{\l}odowska-Curie Actions, European Union; Investissements d'Avenir Labex, Investissements d'Avenir Idex and ANR, France; DFG and AvH Foundation, Germany; Herakleitos, Thales and Aristeia programmes co-financed by EU-ESF and the Greek NSRF, Greece; BSF-NSF and MINERVA, Israel; Norwegian Financial Mechanism 2014-2021, Norway; NCN and NAWA, Poland; La Caixa Banking Foundation, CERCA Programme Generalitat de Catalunya and PROMETEO and GenT Programmes Generalitat Valenciana, Spain; G\"{o}ran Gustafssons Stiftelse, Sweden; The Royal Society and Leverhulme Trust, United Kingdom.
 
In addition, individual members wish to acknowledge support from Chile: Agencia Nacional de Investigaci\'on y Desarrollo (FONDECYT 1190886, FONDECYT 1210400, FONDECYT 1230812, FONDECYT 1230987); China: National Natural Science Foundation of China (NSFC - 12175119, NSFC 12275265, NSFC-12075060); Czech Republic: PRIMUS Research Programme (PRIMUS/21/SCI/017); European Union: European Research Council (ERC - 948254), Horizon 2020 Framework Programme (MUCCA - CHIST-ERA-19-XAI-00), European Union, Future Artificial Intelligence Research (FAIR-NextGenerationEU PE00000013), Italian Center for High Performance Computing, Big Data and Quantum Computing (ICSC, NextGenerationEU), Marie Sklodowska-Curie Actions (EU H2020 MSC IF GRANT NO 101033496); France: Agence Nationale de la Recherche (ANR-20-CE31-0013, ANR-21-CE31-0013, ANR-21-CE31-0022), Investissements d'Avenir Idex (ANR-11-LABX-0012), Investissements d'Avenir Labex (ANR-11-LABX-0012); Germany: Baden-Württemberg Stiftung (BW Stiftung-Postdoc Eliteprogramme), Deutsche Forschungsgemeinschaft (DFG - CR 312/5-1); Italy: Istituto Nazionale di Fisica Nucleare (FELLINI G.A. n. 754496, ICSC, NextGenerationEU); Japan: Japan Society for the Promotion of Science (JSPS KAKENHI JP21H05085, JSPS KAKENHI JP22H01227, JSPS KAKENHI JP22H04944); Netherlands: Netherlands Organisation for Scientific Research (NWO Veni 2020 - VI.Veni.202.179); Norway: Research Council of Norway (RCN-314472); Poland: Polish National Agency for Academic Exchange (PPN/PPO/2020/1/00002/U/00001), Polish National Science Centre (NCN 2021/42/E/ST2/00350, NCN OPUS nr 2022/47/B/ST2/03059, NCN UMO-2019/34/E/ST2/00393, UMO-2020/37/B/ST2/01043, UMO-2021/40/C/ST2/00187); Slovenia: Slovenian Research Agency (ARIS grant J1-3010); Spain: BBVA Foundation (LEO22-1-603), Generalitat Valenciana (Artemisa, FEDER, IDIFEDER/2018/048), La Caixa Banking Foundation (LCF/BQ/PI20/11760025), Ministry of Science and Innovation (MCIN \& NextGenEU PCI2022-135018-2, MICIN \& FEDER PID2021-125273NB, RYC2019-028510-I, RYC2020-030254-I, RYC2021-031273-I, RYC2022-038164-I), PROMETEO and GenT Programmes Generalitat Valenciana (CIDEGENT/2019/023, CIDEGENT/2019/027); Sweden: Swedish Research Council (VR 2018-00482, VR 2022-03845, VR 2022-04683, VR grant 2021-03651), Knut and Alice Wallenberg Foundation (KAW 2017.0100, KAW 2018.0157, KAW 2018.0458, KAW 2019.0447); Switzerland: Swiss National Science Foundation (SNSF - PCEFP2\_194658); United Kingdom: Leverhulme Trust (Leverhulme Trust RPG-2020-004); United States of America: U.S. Department of Energy (ECA DE-AC02-76SF00515), Neubauer Family Foundation.


\clearpage
 
\printbibliography
 
\clearpage
 
\begin{flushleft}
\hypersetup{urlcolor=black}
{\Large The ATLAS Collaboration}

\bigskip

\AtlasOrcid[0000-0002-6665-4934]{G.~Aad}$^\textrm{\scriptsize 102}$,
\AtlasOrcid[0000-0002-5888-2734]{B.~Abbott}$^\textrm{\scriptsize 120}$,
\AtlasOrcid[0000-0002-1002-1652]{K.~Abeling}$^\textrm{\scriptsize 55}$,
\AtlasOrcid[0000-0001-5763-2760]{N.J.~Abicht}$^\textrm{\scriptsize 49}$,
\AtlasOrcid[0000-0002-8496-9294]{S.H.~Abidi}$^\textrm{\scriptsize 29}$,
\AtlasOrcid[0000-0002-9987-2292]{A.~Aboulhorma}$^\textrm{\scriptsize 35e}$,
\AtlasOrcid[0000-0001-5329-6640]{H.~Abramowicz}$^\textrm{\scriptsize 151}$,
\AtlasOrcid[0000-0002-1599-2896]{H.~Abreu}$^\textrm{\scriptsize 150}$,
\AtlasOrcid[0000-0003-0403-3697]{Y.~Abulaiti}$^\textrm{\scriptsize 117}$,
\AtlasOrcid[0000-0002-8588-9157]{B.S.~Acharya}$^\textrm{\scriptsize 69a,69b,m}$
% ...
\bigskip
\\

% ...
$^{29}$Physics Department, Brookhaven National Laboratory, Upton NY; United States of America.\\
% ...
$^{35}$$^{(a)}$Facult\'e des Sciences Ain Chock, R\'eseau Universitaire de Physique des Hautes Energies - Universit\'e Hassan II, Casablanca;$^{(b)}$Facult\'{e} des Sciences, Universit\'{e} Ibn-Tofail, K\'{e}nitra;$^{(c)}$Facult\'e des Sciences Semlalia, Universit\'e Cadi Ayyad, LPHEA-Marrakech;$^{(d)}$LPMR, Facult\'e des Sciences, Universit\'e Mohamed Premier, Oujda;$^{(e)}$Facult\'e des sciences, Universit\'e Mohammed V, Rabat;$^{(f)}$Institute of Applied Physics, Mohammed VI Polytechnic University, Ben Guerir; Morocco.\\
% ...
$^{49}$Fakult\"{a}t Physik , Technische Universit{\"a}t Dortmund, Dortmund; Germany.\\
% ...
$^{55}$II. Physikalisches Institut, Georg-August-Universit\"{a}t G\"ottingen, G\"ottingen; Germany.\\
% ...
$^{69}$$^{(a)}$INFN Gruppo Collegato di Udine, Sezione di Trieste, Udine;$^{(b)}$ICTP, Trieste;$^{(c)}$Dipartimento Politecnico di Ingegneria e Architettura, Universit\`a di Udine, Udine; Italy.\\
% ...
$^{102}$CPPM, Aix-Marseille Universit\'e, CNRS/IN2P3, Marseille; France.\\
% ...
$^{117}$Department of Physics, New York University, New York NY; United States of America.\\
% ...
$^{120}$Homer L. Dodge Department of Physics and Astronomy, University of Oklahoma, Norman OK; United States of America.\\
% ...
$^{150}$Department of Physics, Technion, Israel Institute of Technology, Haifa; Israel.\\
$^{151}$Raymond and Beverly Sackler School of Physics and Astronomy, Tel Aviv University, Tel Aviv; Israel.\\
% ...
$^{a}$ Also Affiliated with an institute covered by a cooperation agreement with CERN.\\
$^{b}$ Also at An-Najah National University, Nablus; Palestine.\\
$^{c}$ Also at Borough of Manhattan Community College, City University of New York, New York NY; United States of America.\\
$^{d}$ Also at Center for High Energy Physics, Peking University; China.\\
$^{e}$ Also at Center for Interdisciplinary Research and Innovation (CIRI-AUTH), Thessaloniki; Greece.\\
$^{f}$ Also at Centro Studi e Ricerche Enrico Fermi; Italy.\\
% ...
$^{m}$ Also at Department of Physics, King's College London, London; United Kingdom.\\
% ...

\end{flushleft}


\end{document}